\documentclass[prb,aps,showpacs,twocolumn,floatfix,superscriptaddress]{revtex4}
\usepackage{eurosym}
\usepackage{amsmath,amsfonts,amssymb,graphics,graphicx,color,bm}
\usepackage{hyperref}
\usepackage[percent]{overpic}

\begin{document}

\title{Chiral phases of two-dimensional hard-core bosons with frustrated ring-exchange}

\author{Daniel Huerga}
\affiliation{Instituto de Estructura de la Materia, C.S.I.C., Serrano 123, E-28006 Madrid, Spain}
\author{Jorge Dukelsky}
\affiliation{Instituto de Estructura de la Materia, C.S.I.C., Serrano 123, E-28006 Madrid, Spain}
\author{Nicolas Laflorencie}
\affiliation{Laboratoire de Physique Th\'eorique, IRSAMC, Universit\'e de Toulouse, CNRS, 31062 Toulouse, France}
\author{Gerardo Ortiz}
\affiliation{Department of Physics, Indiana University, Bloomington, IN 47405, USA}

\begin{abstract}
We study the zero temperature phase diagram of two-dimensional hard-core bosons on a 
square lattice with nearest neighbour and plaquette (ring-exchange) hoppings, at arbitrary densities, 
by means of a hierarchical mean-field theory. In the frustrated regime, 
where quantum Monte Carlo suffers from a sign problem, we find a rich phase diagram where exotic 
states with nonzero chirality emerge. Among them, novel insulating phases, characterized by 
nonzero bond-chirality and plaquette order, are found over a large region of the parameter space. 
In the unfrustrated regime, the hierarchical mean-field approach improves over the standard 
mean-field treatment as it is able to capture the transition from a superfluid to a valence bond 
state upon increasing the strength of the ring-exchange term,  in qualitative agreement with 
quantum Monte Carlo results.
\end{abstract}

\pacs{
05.30.Jp, 
75.10.-b, 
75.10.Jm, 
64.70.Tg  
}
\maketitle

\section{\label{intro}Introduction}

In quantum systems multi-particle exchange competing interactions often play an important role in establishing complex thermodynamic phases with unconventional orders \cite{roger_rev}. Those interactions are known to be relevant in certain bosonic and fermionic systems, such as solid $^4$He and $^3$He \cite{roger}. In particular, four-spin ring exchange processes  have been argued to be necessary in explaining magnetic excitations in cuprate high-$T_c$ superconductors \cite{coldea}. Moreover, others claim that they can be essential in understanding the pseudogap phase in the cuprates. 

\begin{figure}[t]
\begin{overpic}[angle=0, width=0.95\columnwidth, height=6cm]{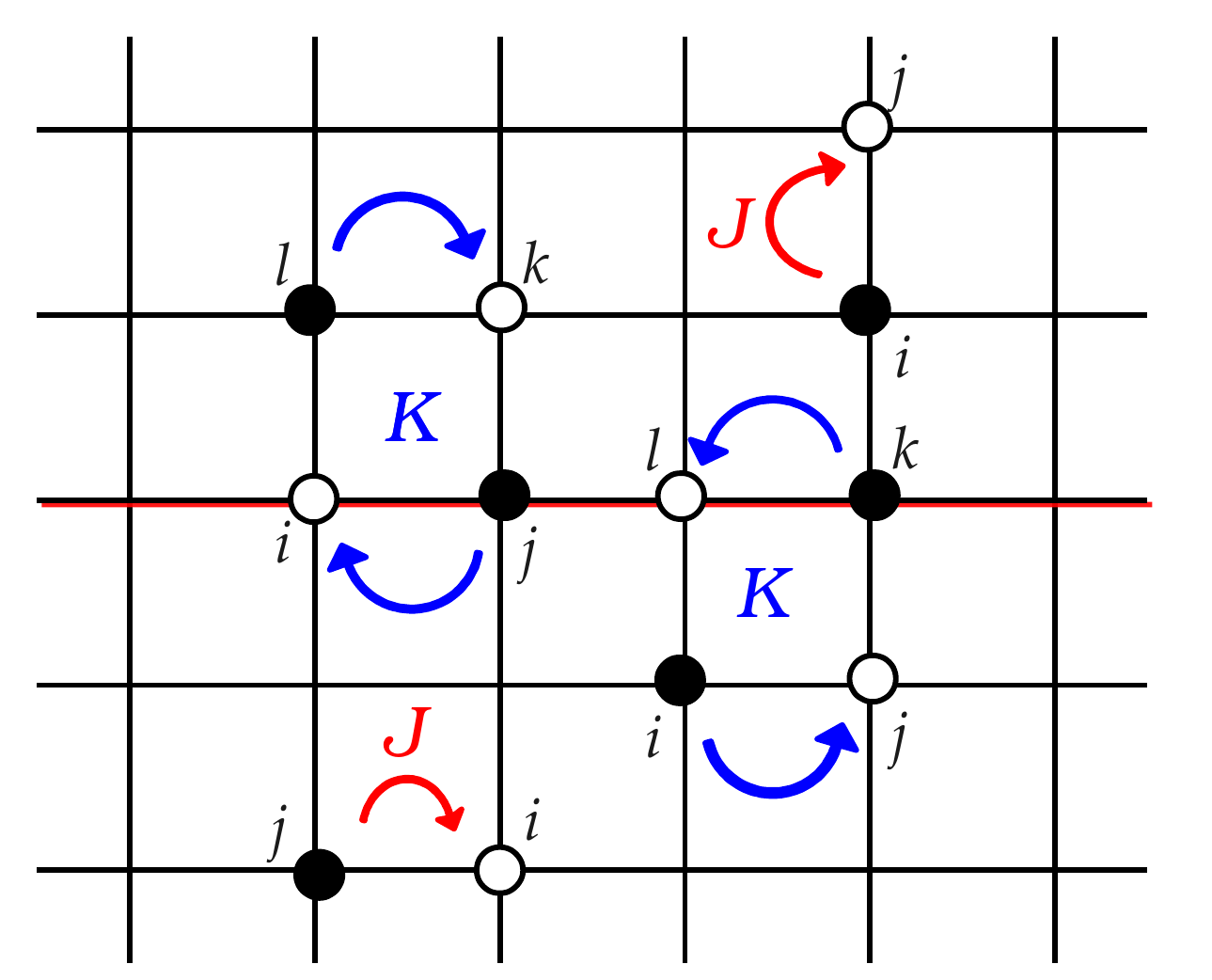}
\put (4,38) {\large$\Gamma_{\nu}$}
\end{overpic}
\caption{\label{JKscheme}
(Color online) Graphical representation of the interactions terms in the $J$-$K$ Hamiltonian. Filled circles stand for initial state and empty circles for the final state after a single-boson hopping of magnitude $J$ or ring-exchange process. The ring-exchange is a two-boson hopping from opposite corners of a plaquette to the other two with constant coupling $K$. This latter process preserves the total number of bosons in every line and column of the lattice (see text). In the frustrated regime $(K>0)$, it favors bond-chiral order.}
\end{figure}

Different kinds of ring-exchange interactions have been proposed in the literature. In the present manuscript we are interested in a particular ring-exchange process competing with a single-particle kinetic energy term. We investigate the quantum phase diagram of the so-called $J$-$K$ model defined by \cite{qmc_hf,qmc}
\begin{eqnarray}
H&=& -J\sum_{\langle ij\rangle}\hat{B}_{ij}+ K\sum_{\langle ijkl\rangle}\hat{P}_{ijkl}
-\mu\sum_{j}n_{j},
\label{JK}
\end{eqnarray}
where $n_{j}=a^{\dag}_{j}a_{j}^{\;}$ is the density operator, and
\begin{eqnarray}
\hat{B}_{ij}&=& a^{\dag}_{i}a_{j}^{\;} + a^{\dag}_{j}a_{i}^{\;},\label{B}\\
\hat{P}_{ijkl}&=& a^{\dag}_{i}a^{\dag}_{k}a_{j}^{\;}a_{l}^{\;} + a^{\dag}_{l}a^{\dag}_{j}a_{k}^{\;}a_{i}^{\;},\label{P}
\end{eqnarray}
are the hopping and plaquette operators written in terms of creation, $a^{\dag}_{j}$, and annihilation, $a_j^{\;}$, hard-core boson operators at site $j$ of a square lattice with $L_x\times L_y$ sites. The nearest-neighbor hopping amplitude is $J>0$, $\mu$ is the chemical potential controlling the density of the system, and $K$ is the strength of the ring-exchange process where two hard-core bosons on (diagonally) opposite corners of a plaquette $\langle ijkl\rangle$ hop simultaneously to the other two corners, as schematically represented in Fig. \ref{JKscheme}.

The $J$-$K$ model (\ref{JK}) can be equivalently written as an (easy-plane) XY model with a four-spin interaction, via the Matsubara-Matsuda transformation \cite{mm}. By virtue of this mapping, creation (annihilation) operators of hard-core bosons are simply replaced by ladder operators of the $SU(2)$ algebra in the $S=1/2$ representation, $a^{\dag}_{j}= S^{+}_{j}$ and $a_{j}^{\;}= S^{-}_{j}$, while the number operator is replaced by the Cartan operator, $n_{j}= S^{z}_{j}+1/2$. In terms of these $S=1/2$ spin operators, the Hamiltonian $(\ref{JK})$ can be rewritten as follows 
\begin{eqnarray}
H &=& 
- 2  J\sum_{\left\langle ij \right\rangle} \left( S^{x}_{i}S^{x}_{j} + S^{y}_{i}S^{y}_{j} \right)
- \mu  \sum_{j} \left( S^{z}_{j} + \frac{1}{2} \right) \nonumber\\
&& + 2 K  \sum_{ \left\langle ijkl \right\rangle }
(
S^{x}_{i}S^{x}_{j}S^{x}_{k}S^{x}_{l}
+ S^{y}_{i}S^{y}_{j}S^{y}_{k}S^{y}_{l} \notag\\
&&~~~~~~~~~~~~~
+ S^{x}_{i}S^{x}_{j}S^{y}_{k}S^{y}_{l}
+ S^{y}_{i}S^{y}_{j}S^{x}_{k}S^{x}_{l} \notag\\
&&~~~~~~~~~~~~~
+ S^{y}_{i}S^{x}_{j}S^{x}_{k}S^{y}_{l} 
+ S^{x}_{i}S^{y}_{j}S^{y}_{k}S^{x}_{l} \notag\\
&&~~~~~~~~~~~~~
- S^{x}_{i}S^{y}_{j}S^{x}_{k}S^{y}_{l} 
- S^{y}_{i}S^{x}_{j}S^{y}_{k}S^{x}_{l}
),\label{Hspin}
\end{eqnarray}
where $S^x_j=(S^+_j +S^-_j)/2$ and $S^y_j=(S^+_j - S^-_j)/2{\mathrm i}$. 

The $J$-$K$ model is not $SU(2)$ invariant, as it is the case of the $J$-$Q$\cite{sandvikJQ,isaevJQ} and related ring-exchange models\cite{Laeuchli05}, but displays a lower global $U(1)$ symmetry. Moreover, for $J$=0, the $J$-$K$ model has $d=1$ $U(1)$ gauge-like symmetries \cite{zoharortiz}, a total of $L_x+L_y$ unitary operators 
\begin{eqnarray}
\hat{\cal O}_{\nu}&=&e^{ {\mathrm{i}} \phi \sum_{j\in \Gamma_\nu} n_j}
\label{colrowsym}
\end{eqnarray}
where $\Gamma_\nu$ represents any horizontal or vertical line of the lattice, of length $L_x$ or $L_y$, respectively (see Fig. \ref{JKscheme}). These $d=1$ symmetries, leading to dimensional reduction \cite{zoharortiz}, constrain the dynamics of the model, as already indicated for a soft-core bosonic version in Ref. [\onlinecite{paramekanti}], 
and leads to stripe-like correlations. This $K$-only model
\begin{eqnarray}
H_K&=& K\sum_{\langle ijkl\rangle}\hat{P}_{ijkl},
\label{Konly}
\end{eqnarray}
also displays a chiral symmetry, with a unitary operator
\begin{eqnarray}
{\cal C}&=& e^{{\mathrm{i}} \frac{\pi}{2} \sum_{j \in A} n_j}
\label{Konlychiral}
\end{eqnarray}
that anti-commutes with $H_K$, and where the sum is performed over sites $j$ of one of the disjoint sublattices $A$ of the original bipartite lattice. This, in turn, implies that the  eigenvalue spectrum of $H_K$ is symmetric around zero with the operator ${\cal C}$ connecting the ground state of $H_K$ with that of $H_{-K}$, i.e.,
\begin{eqnarray}
| \Psi_0(-K)\rangle&=& {\cal C} | \Psi_0(K)\rangle .
\end{eqnarray}
This means that correlation functions involving density operators are trivially related. For example,
\begin{eqnarray}
\hspace*{-0.4cm}
\langle \Psi_0(K)|n_in_j| \Psi_0(K)\rangle&=& \langle \Psi_0(-K)|n_in_j| \Psi_0(-K)\rangle,
\end{eqnarray}
with the remarkable consequence that long-range order in any density correlation function is 
independent of the sign of $K$. 
One can show that the Hamiltonian $H_K$ has a zero energy eigenspace that can be exactly determined by all those tilings of the lattice with plaquette configurations that exclude the two (out of sixteen) involving only two particles occupying opposite sites of a diagonal. This eigenspace is massively degenerate.

It is interesting to remark that $H_K$ is invariant under transmutation of exchange statistics. This means that one can write $H_K$ in terms of hard-core {\it anyons} \cite{adinphy} (which includes spinless fermions when the statistical angle is $\pi$) and the resulting eigenspectrum remains invariant. The origin of this invariance is, precisely, the existence of the $d$=1 gauge-like symmetries mentioned above. 

For $K>0$, the ring exchange term dynamically frustrates the usual hopping $J$. This fact is at the root of the sign problem encountered in quantum Monte Carlo (QMC) simulations of the model. The $J$-$K$ model has been studied by QMC techniques in the unfrustrated region ($K<0$), at half filling ($\mu$=0) \cite{qmc_hf} and away from half filling \cite{qmc}. These studies have been motivated by the proposal of a new gapless Bose liquid phase dubbed exciton Bose liquid \cite{paramekanti}. In addition, the frustrated region ($K>0$) has been explored at half filling by a semiclassical approximation \cite{schaffer} revealing the emergence of a bond-chiral superfluid (CSF) phase at $K=2$ characterized by nonvanishing condensate and superfluid densities and a nonzero bond-chirality. 

In the present work we determine the quantum phase diagram of the $J$-$K$ model (\ref{JK}) in the frustrated regime of the ring-exchange interaction $(K>0)$ by means of the hierarchical mean-field theory (HMFT) \cite{hmft,isaevJQ}. The HMFT is a useful tool to unveil strongly correlated phases of matter where other methods face significant problems or are even inapplicable. The method is based on the identification of the relevant elementary degrees of freedom which capture the necessary quantum correlations in order to describe the essential features of the phases present in the system under study. The set of operators which describe the quantum states of the new degrees of freedom and their algebra provide the {\it hierarchical language} \cite{adinphy} adequate to describe the system. The use of this method combined with bond-algebra techniques and duality mappings \cite{bond1,duality1} makes the HMFT a suitable and powerful technique to investigate phase diagrams of strongly correlated systems. 

%
\begin{figure*}[t]
\includegraphics[angle=0,clip=true,scale=0.42]{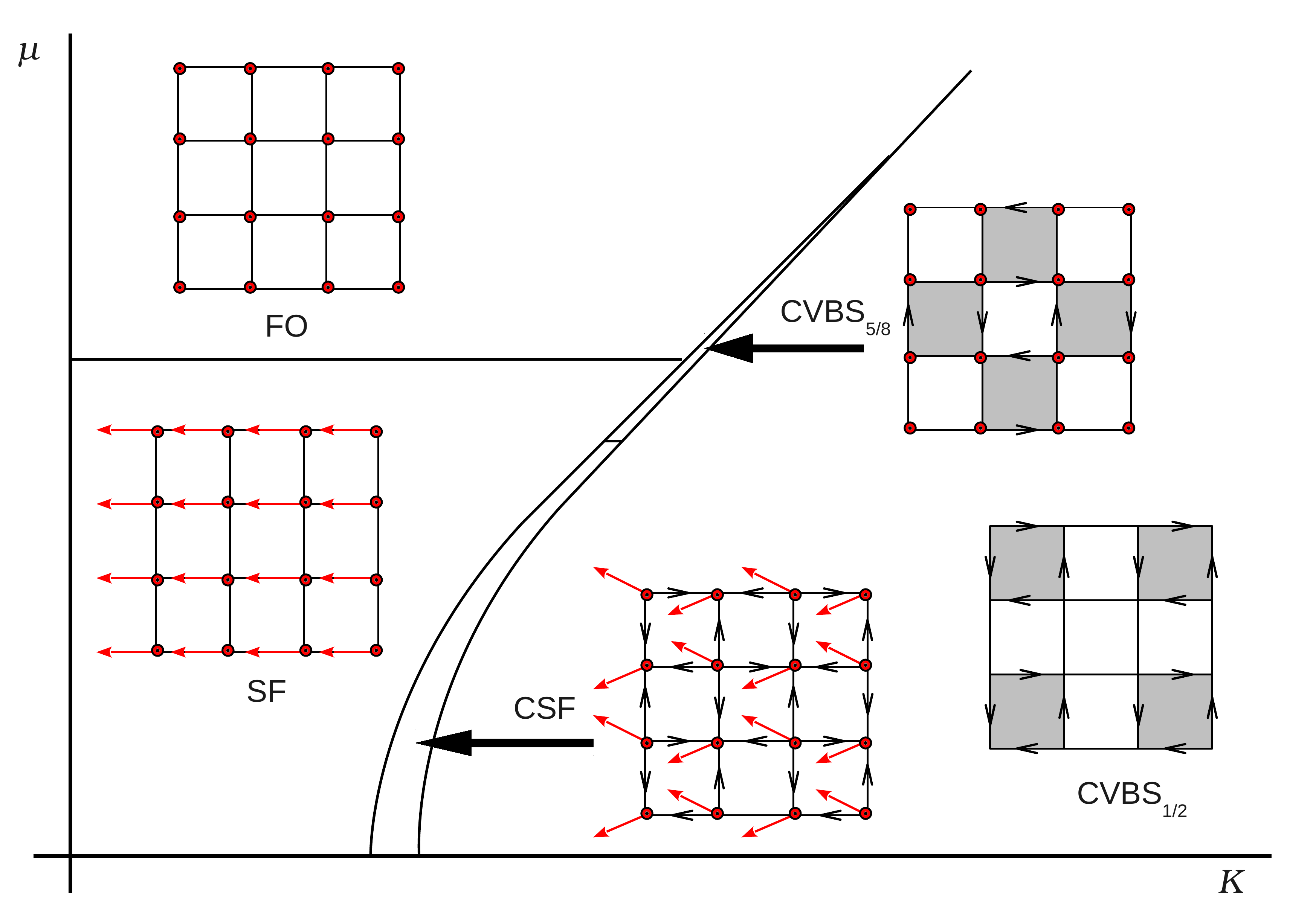}
\caption{\label{phases} (Color online) Schematic representation of the superfluid and solid phases obtained by means of the HMFT--4$\times$4 for the $J$-$K$ model (\ref{JK}) in the frustrated regime $(K\geq 0)$ and with an external chemical potential $(\mu\geq 0)$. Phases are pictorially represented in the spin language: filled red arrows indicate the $x$-$y$ projection of the expectation value of the spin operator $\langle \mathbf{S}\rangle$ at each site, while dots indicate the projection along the $z$-axis. Empty black arrows along the bonds of the lattice indicate the bond-chirality (see text). The bosonic language is utilized to name the phases: superfluid (SF), fully occupied (FO), bond-chiral superfluid (CSF), half filled valence bond-chiral solid (CVBS$_{1/2}$) and valence bond-chiral solid of density $\rho=5/8$ (CVBS$_{5/8}$). Shaded and white plaquettes correspond to the alternating strength pattern of the plaquette operator characteristic of the CVBS$_{\rho}$ phases (see text). The phase diagram is symmetric with respect to the $\mu=0$ line due to particle-hole symmetry. Under this symmetry, the FO region transforms onto the vacuum of hard-core bosons.}
\end{figure*}
In practice, we carry out this program by tiling the original lattice into clusters of equal size and shape $(L\times L)$ preserving most of the symmetries. Short-range quantum correlations within each cluster are exactly computed by representing each state of the cluster as the action of a new creation bosonic operator over the vacuum of an enlarged Fock space. The mapping which relates the original and the new bosonic operators can be considered as an extension of the Schwinger-boson mapping \cite{auerbach} of $S=1/2$ spin operators in the hard-core language \cite{hmft} , or an extension of the slave-boson mapping of canonical bosons\cite{dickerscheid}, to clusters. These new set of cluster bosonic operators, dubbed {\it composite bosons} (CB)\cite{cbmft}, carry a quantum label which corresponds to the state of the cluster which they describe. As a consequence, the physical subspace of this new enlarged cluster space is defined by those states having only one CB on each superlattice site. As the original operators and the new CB operators are related by a canonical mapping, the Hamiltonian can be re-expressed in terms of the new language and treated by means of standard many-body techniques. As a first approximation, we use a product wave function, that we call Gutzwiller wave function. Other cluster mean-field methods have been shown to be equivalent to this approximation\cite{danshita}. 

In the present manuscript, we use clusters of size $1\times 1,~2\times2~\text{and}~4\times 4$. Within the Gutzwiller wave function approach, the HMFT--1$\times$1 turns out to be equivalent to a classical approximation, so that a linear spin-wave dispersion over the superfluid and the two-sublattice bond-chiral superfluid are easily computed. For clusters larger than a single site, HMFT--$L$$\times$$L$ allows for the existence of solid phases with bond and plaquette orders which cannot be accounted for by the classical approximation.

We determine the quantum phase diagram on the $(K,\mu)$ plane, assuming that all energies are given in units of the hopping parameter $J$. We obtain various superfluid and solid phases, some of them characterized by the presence of bond-chiral order. In the frustrated regime ($K>0$), we find a conventional uniform superfluid (SF) and fully occupied (FO) or empty (VAC) phases, as well as a less conventional bond-chiral superfluid (CSF) and two novel insulating valence bond-chiral solid phases (CVBS$_{\rho}$) at densities $\rho=1/2$ and $\rho=5/8$. The latter are characterized by an alternating pattern of the expectation values of the hopping (\ref{B}), plaquette (\ref{P}), and bond-chiral operators defined below. Contrary to other chiral fluid or solid phases \cite{hassanieh,zaletel}, the bond-chiral phases encountered here do not develop spontaneous loop currents. Instead, they form {\it source-and-drain} patterns, as it is shown schematically in Fig. \ref{phases}. Notice that HMFT leads to an explicit breaking of translational symmetry, which should be restored in the thermodynamic limit. Therefore, one cannot draw rigorous conclusions on the order of the phase transitions based solely on a fixed coarse graining. One can remedy this situation by performing finite-size scaling of the HMFT cluster. As the size of the cluster simulated gets larger we get closer to the exact solution in the thermodynamic limit. It is remarkable that a single wave function allow us to map the full phase diagram, thus containing information about various competing orders.  

Studying how quantum phases evolve as the size of the clusters increases allows us to assess the stability of the solution obtained in the previous steps. As an example, the stability of the CSF phase obtained within the classical approximation reduces to a region between the uniform superfluid and the new half filled CVBS$_{1/2}$ phase when computed with clusters of size 2$\times$2. Moreover, a novel CVBS$_{5/8}$ phase of density $\rho=5/8$ emerges when clusters of size 4$\times$4 are utilized, thus reducing the region of stability of the CSF phase. We cannot rule out  the appearance of new commensurate CVBS$_\rho$ phases, with even larger characteristic correlation length, when larger clusters are used. 

The current control and manipulation of cold atom systems in optical lattices allow experimentalists to simulate and probe a variety of condensed matter lattice Hamiltonians whith unprecedented accuracy. Two recent theoretical proposals \cite{coldat,rydberg} suggest ways to implement the ring-exchange  Hamiltonian (\ref{JK}) in optical lattices. The experimental realization of these proposals could test the existence of the chiral phases obtained in the present work.

The outline of the paper is as follows. In Sec. \ref{ClasPhaseDiag} we compute the classical phase diagram obtaining three phases: FO, SF and CSF. Bond-chirality emerges from the fact  that the ring-exchange  interaction is frustrating $(K>0)$. In Sec. \ref{HMFT} we present the CB mapping which relates the original bosonic hard-core operators to a new set of operators representing cluster states. By means of this mapping we re-express the $J$-$K$ Hamiltonian (\ref{JK}) in a new language. This CB Hamiltonian encodes the complete information of the original $J$-$K$ model in the definition of certain matrix elements, whose details are provided  in Appendix \ref{matrixel}. We then apply the Gutzwiller approximation and show that using one-site clusters is equivalent to the classical approximation and compute the spin-wave excitations in Appendix \ref{LSWT}. For larger clusters, we show that the Gutzwiller approximation is equivalent to the exact diagonalization of a finite cluster embedded in a self-consistently defined environment (Sec. \ref{gutz}). We define the relevant order parameters and observables needed to characterize the quantum phases in Sec.\ref{OPs}. In Sec. \ref{HMFphasediag} we present the quantum phase diagram within the HMFT--2$\times$2 and HMFT--4$\times$4 schemes. Finally, we summarize the main results in Sec.\ref{conclusions}.

\section{\label{ClasPhaseDiag}Classical phase diagram}

\subsection{Classical approximation}
As a first approach to the phase diagram of the Hamiltonian $(\ref{JK})$, we study in this section the ground state phases within the classical limit. In this limit, the $SU(2)$ spin operator ${\bf S}_j=\left(S^x_j,S^y_j,S^z_j\right)$ can be approximated by a classical spin vector, that is, $\vec{S}_{j}=S \left(\sin\theta_{j}\cos\phi_{j},\sin\theta_{j}\sin\phi_{j},\cos\theta_{j}\right)$. Applying this approximation to the Hamiltonian $(\ref{Hspin})$, the classical energy 
(having fixed $J=1$) is a function of the classical spin angles $\{\theta_{j},\phi_{j}\}$,
\begin{eqnarray}
\mathcal{E}&=& -2S^{2}\sum_{\left\langle ij \right\rangle} \sin\theta_{i}\sin\theta_{j}\cos(\phi_{i}-\phi_{j})\nonumber\\
&&+ 2KS^{4} \sum_{\left\langle ijkl \right\rangle}
\sin\theta_{i}\sin\theta_{j}\sin\theta_{k}\sin\theta_{l}\nonumber\\
&& \times \cos(\phi_{i}-\phi_{j}+\phi_{k}-\phi_{l})\nonumber\\
&&- \mu \sum_{j} \left( S\cos\theta_{j} +\frac{1}{2} \right).
\label{e_clas}
\end{eqnarray}
For $S=1/2$, the case of interest here, the energy $(\ref{e_clas})$ can be equivalently obtained by taking the expectation value of Hamiltonian $(\ref{JK})$ with a product wave function in which the spins of the lattice are in a Bloch sphere representation
\cite{sakurai},
\begin{equation}
\vert \Psi_0 \rangle = 
\prod_{j} 
\left[\sin\left(\frac{\theta_{j}}{2}\right)e^{{\mathrm i}\frac{\phi_{j}}{2}} \lvert \downarrow\rangle
+\cos\left(\frac{\theta_{j}}{2}\right)e^{-{\mathrm i}\frac{\phi_{j}}{2}}\lvert \uparrow\rangle \right].\label{prod_wf}
\end{equation}
By virtue of the Matsubara-Matsuda mapping, the bosonic counterpart is straightforwardly obtained by replacing $\lvert \uparrow \rangle\rightarrow \lvert 1 \rangle$ and $\lvert \downarrow \rangle\rightarrow \lvert 0 \rangle$. Therefore, minimizing expression $(\ref{e_clas})$ with respect to the variational parameters $\{\theta_{j},\phi_{j}\}$ leads to the classical solution or, equivalently, to a variational approximation with the trial wave function $(\ref{prod_wf})$. We assume a trial two-sublattice product wave function where the two sublattices, $A$ and $B$, form a checkerboard structure. Within this approximation, the variational ansatz $(\ref{prod_wf})$ has four variational parameters $(\theta_{A},\phi_{A},\theta_{B},\phi_{B})$. However, by fixing a global phase we can choose $\phi_{A}=-\phi_{B}=\phi$ without loss of generality. This ansatz is able to describe a wide range of two-sublattice bosonic phases, namely: charge density-wave (CDW) with $\mathbf{q}=(\pi,\pi)$ ordering wave vector, checkerboard supersolid (CSS) and bond-chiral superfluid (CSF); apart from the uniform ones: superfluid (SF) and fully occupied (FO) or empty (VAC). Their semiclassical wave functions are characterized by
%
\begin{figure}[t]
\includegraphics[clip=true,width=1.\columnwidth]{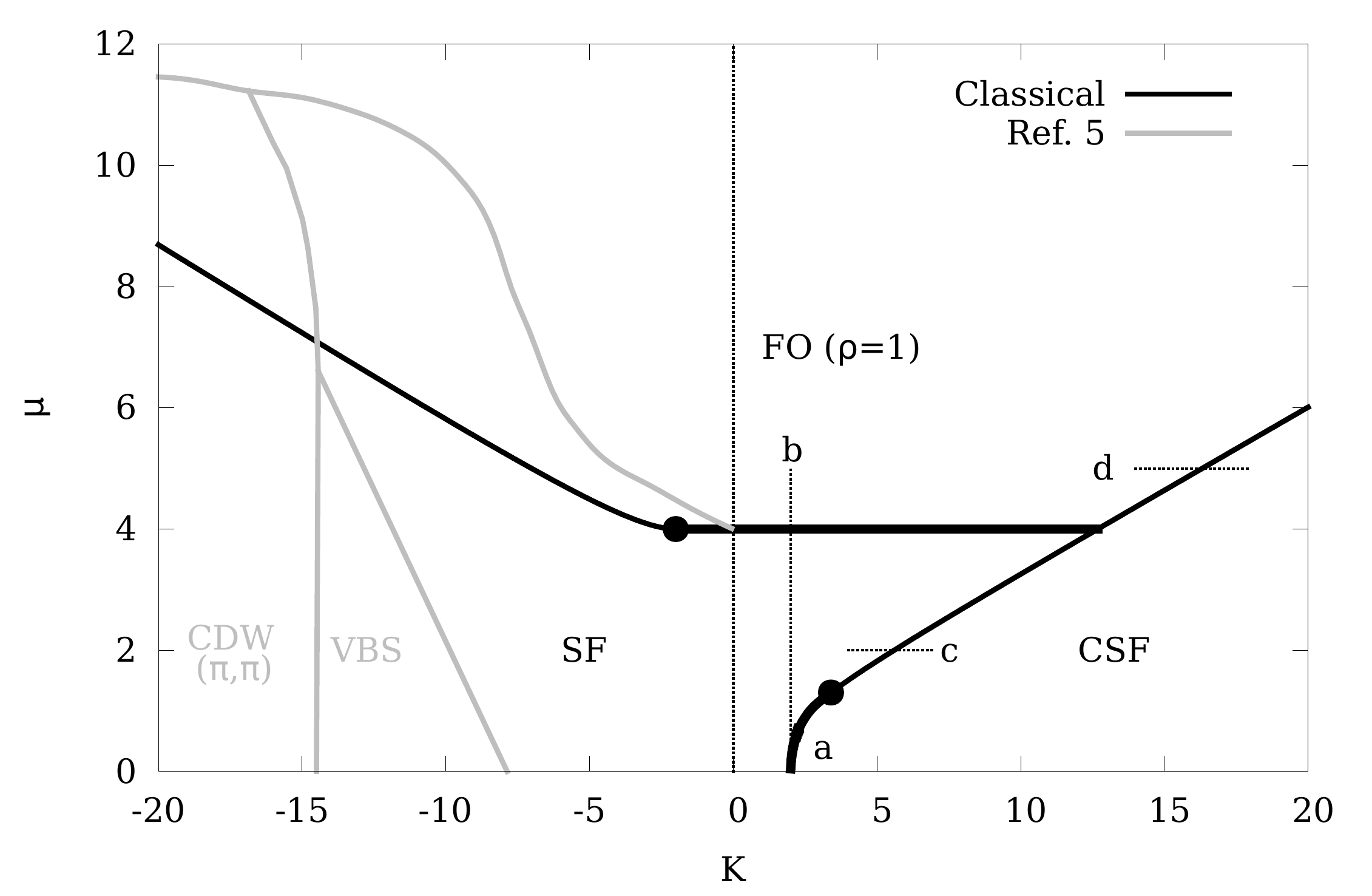}
\caption{Classical phase diagram (black) and schematic phase diagram obtained by means of QMC ($K<0$) from Ref. \onlinecite{qmc} (gray). The dashed line at $K=0$ marks the division between the unfrustrated $(K<0)$ and the frustrated region $(K> 0)$, where QMC is not applicable. Thin (thick) solid black lines correspond to first (second) order phase transitions. Four cuts ($a,b,c$ and $d$) across the phase transitions are also indicated with dashed lines. The filled (black) circles indicate potential tricritical points (TCP).}
\label{CLdiag}
\end{figure}
\begin{eqnarray}
\text{FO}&:& \theta_{A}=\theta_{B}=0;\label{FO}\\
\text{VAC}&:& \theta_{A}=\theta_{B}=\pi;\label{VAC}\\
\text{SF}&:& 0<\theta_{A}=\theta_{B}< \pi,~ \phi=0;\label{sf}\\
\text{CSF}&:&  0<\theta_{A}=\theta_{B}< \pi,~ 0<\phi< \pi/2;\label{csf}\\
\text{CDW}&:& \theta_{A}=0,~\theta_{B}=\pi;\label{neel}\\
\text{CSS}&:& \theta_{A}\neq\theta_{B},~\theta_{A}\neq 0,\pi,~\theta_{B}\neq 0,\pi.\label{css}
\end{eqnarray}
In terms of spins, the FO phase of hard-core bosons corresponds to a fully polarized ferromagnet. The SF phase is characterized by the Bose-Einstein condensation (BEC) of hard-core bosons at momentum $\mathbf{k}=(0,0)$, which breaks the global $U(1)$ symmetry of the Hamiltonian (\ref{JK}). It corresponds to a ferromagnet with nonzero projection over the $x\text{-}y$ plane and nonzero spin stiffness. The CSF phase, characterized by nonzero bond-chirality and a two-component BEC at $\mathbf{k}=(0,0)$ and $(\pi,\pi)$, corresponds to a canted magnet with staggered azimuth orientation of the spins $(\phi)$. The $(\pi,\pi)$ CDW, corresponds to the N\'eel phase in which ``up'' and ``down'' spins alternate forming a checkerboard pattern. The CSS, characterized by the coexistence of $(\pi,\pi)$ CDW order and BEC at $\mathbf{k}=(0,0)$, corresponds to a staggered magnet with two sublattices having different projections over the $z$ axis. 

Substituting $S=1/2$ in $(\ref{e_clas})$, the classical energy takes the form
\begin{eqnarray}
\mathcal{E}/N&=& - \sin\theta_{A}\sin\theta_{B}\cos(2\phi)\nonumber\\
&&+ \frac{K}{8} 
\sin^{2}\theta_{A}\sin^{2}\theta_{B}\cos(4\phi)\nonumber\\
&&- \frac{\mu}{4} \left( \cos\theta_{A} + \cos\theta_{B}+2 \right),
\label{e_var}
\end{eqnarray}
where $N$ is the number of sites of a square lattice with periodic boundary conditions (PBC). Minimization of $(\ref{e_var})$ with respect to the angle parameters gives rise to three of the five phases described above (\ref{FO})-(\ref{css}), depending on the region of the parameter space $(K,\mu)$: FO, SF and CSF. In the three cases, the ground state wave functions satisfy $\theta_{A}=\theta_{B}=\theta$. Both SF and CSF display phase coherence, i.e. a rigid phase $\phi$ which is either constant $\phi=0$ in the SF state or staggered ($\phi=\phi_A=-\phi_B$) in the CSF, where it satisfies
\begin{equation}
\cos\left(2\phi\right)=\frac{2}{K\sin^{2}\theta}.
\label{eq:phi}
\end{equation}
%
\begin{figure}[t]
\includegraphics[clip=true,width=\columnwidth]{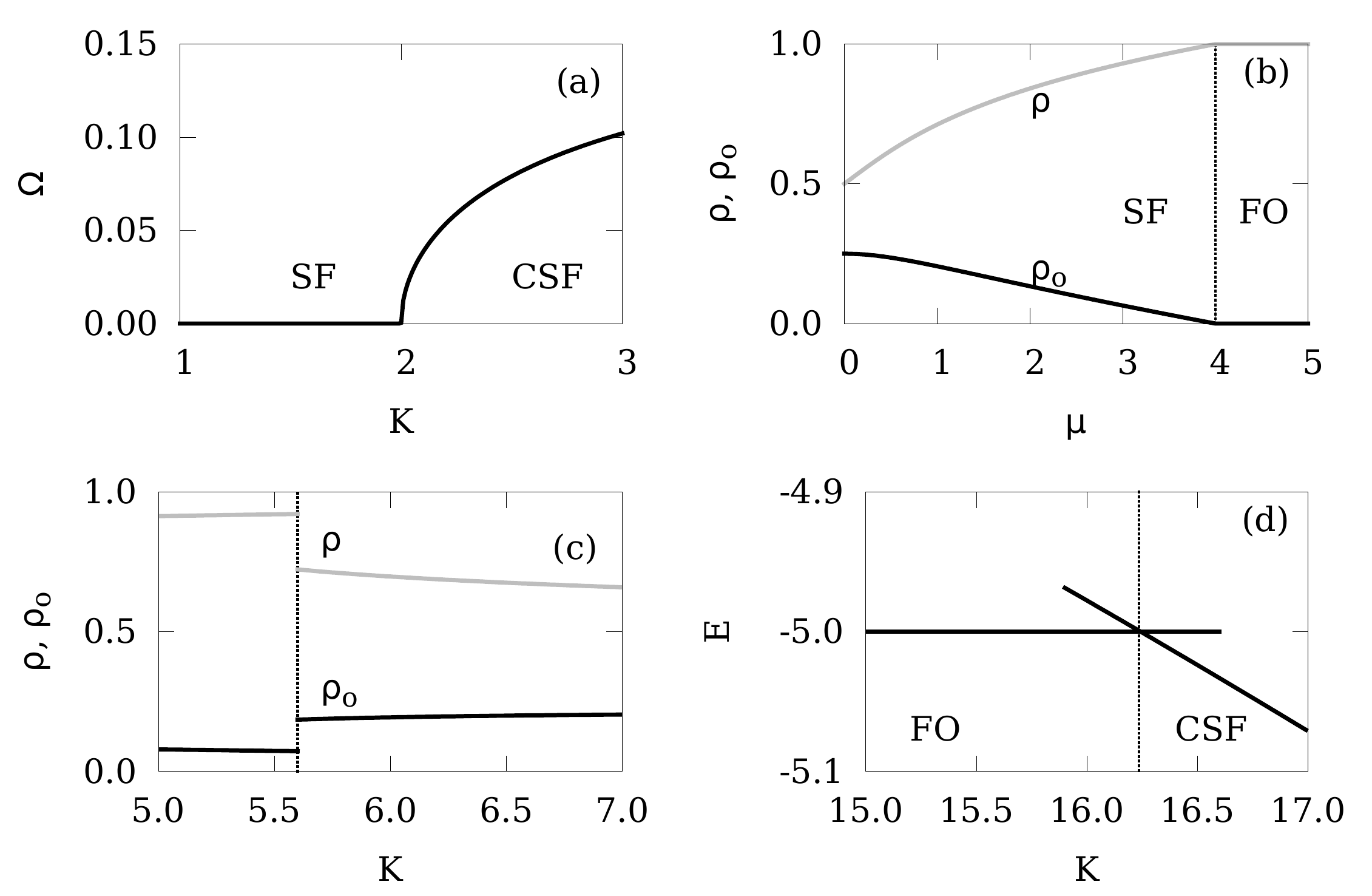} 
\caption{
Energy (E), bond-chiral OP ($\Omega$), condensate density ($\rho_o$) and total density ($\rho$) across four cuts in the classical phase diagram (Fig. $\ref{CLdiag}$). Panel (a): Bond-chiral OP across the SF-CSF transition at $\mu=0$. Panel (b): Total density and condensate density for $K=2$. Panel (c): Total density and condensate density for the CSF (black) and SF (gray) solutions along $\mu=2$. The dashed line at $K=5.6$ indicates the point at which the first order transition occurs. Panel (d): Energy crossing of the FO and CSF phases along $\mu=5$.}
\label{CLcuts}
\end{figure}
\subsection{Order parameters}
%
To characterize these phases, we compute two different order parameters (OPs): the condensate density associated to a bosonic superfluid and the bond-chiral OP. 

The condensate density is derived from the single-particle density matrix, i.e., $\rho_{ij}=\langle\Psi_0\vert a^{\dag}_i a_j\vert\Psi_0\rangle$, which, for a translational invariant system, is diagonal in momentum space
\begin{equation}
\rho_{\mathbf{k}}=\frac{1}{N^{2}}\sum_{ij}e^{-{\mathrm{i}}\mathbf{k}(\mathbf{r}_i - \mathbf{r}_j)}
\rho_{ij}.\label{cond_dens}
\end{equation} 
In the thermodynamic limit, a macroscopic eigenvalue of the single-particle density matrix signals the onset of BEC and defines the condensate density. Within the SF phase, we find a unique macroscopic eigenvalue, at momentum $\mathbf{k}=(0,0)$,
\begin{equation}
\rho_{0}=\frac{1}{4}\sin^{2}\theta,\label{def_rho}
\end{equation}
whereas a second macroscopic eigenvalue appears at $\mathbf{k}=(\pi,\pi)$ within the CSF phase. In this case, the condensate density has two components given by,
\begin{equation}
\rho_{0}=\frac{1}{4}\sin^{2}\theta\cos^{2}\phi
\end{equation}
and
\begin{equation}
\rho_{\pi}=\frac{1}{4}\sin^{2}\theta\sin^2\phi.
\end{equation}
Notice that the CSF phase displays BEC fragmentation, although the uniform component ($\mathbf{k}=(0,0)$) remains dominant over the staggered one ($\mathbf{k}=(\pi,\pi)$) at any finite $K$ with $\rho_0=\rho_\pi + (2K)^{-1}$ for $2\le K<\infty$. However, such a BEC fragmentation observed within the classical treatment is not expected to survive to interactions and quantum fluctuations\cite{nozieres}. 

The bond-chiral operator is defined as the $z$-component of the vector chirality, i.e., $\hat{\Omega}_{ij}= \left(\mathbf{S}_{i}\times\mathbf{S}_{j}\right)_{z}$ \cite{schaffer}, which can be written in the bosonic language as
\begin{equation}
\hat{\Omega}_{ij}=\frac{\mathrm{i}}{2}\left( a^{\dagger}_{i}a_{j} - a^{\dagger}_{j}a_{i}\right),\label{bo_clas}
\end{equation}
where $\mathrm{i}$ stands for the imaginary unit and $i,j$ are two nearest neighbour sites. The bond-chiral operator $(\ref{bo_clas})$ is proportional to the current density of charged bosons, which is defined as $\mathbf{I}_{ij}={\mathrm i}\left( a^{\dag}_{i}a_{j}-a^{\dag}_{j}a_{i} \right) (q/\hbar)\hat{\mathbf{r}}_{ij}$ \cite{hassanieh}, where $q$ is the charge of a boson and $\hat{\mathbf{r}}_{ij}= \left(\mathbf{r}_{j}-\mathbf{r}_{i}\right)/\vert \mathbf{r}_{j}-\mathbf{r}_{i}\vert$. We define the bond-chiral OP as
\begin{equation}
\Omega=\frac{1}{N_{b}}\sum_{\langle ij\rangle} \vert \langle \hat{\Omega}_{ij} \rangle \vert,\label{bchop}
\end{equation}
where $N_{b}=2N$ is the total number of bonds on the square lattice. Computing (\ref{bchop}) with the CSF wave function (\ref{csf}) we find
\begin{equation}
\Omega= \frac{1}{4}\sin\theta\sin(2\phi).
\end{equation}
Differently from other chiral superfluids \cite{zaletel}, the CSF does not present spontaneous currents around closed loops in the lattice. On the contrary, the system forms a checkerboard pattern of {\it source-and-drain} sites, as it is schematically represented in Fig. $\ref{phases}$.
\subsection{Phase diagram}
Fig. \ref{CLdiag} shows the classical phase diagram obtained by minimizing the classical energy $(\ref{e_var})$ in the parameter space $(K,\mu)$. The phases are characterized by the OPs introduced above. First order phase transitions take place at an energy crossing of two different trial wave functions resulting in a discontinuity of the OPs. Second order phase transitions are determined at those points in the parameter space where the OPs vanish continuously. Also displayed in this figure is the schematic phase diagram derived from QMC results in Ref.[\onlinecite{qmc}] for the unfrustrated region $(K<0)$; the frustrated region $(K>0)$ is problematic for QMC due to the ``sign problem''. For $K<0$ we find two phases, the FO and the SF. The VBS cannot be obtained within the single site product wave function approximation $(\ref{prod_wf})$. We find a saddle point of the variational energy for the $(\pi,\pi)$ CDW wave function (\ref{neel}), however it possesses higher energy than the SF solution. For $K>0$ we find three different phases: FO, SF, and CSF. At half filling ($\mu=0$) and up to $\mu\simeq 1.3$ the transition from SF to CSF is of second order type, while it is first order for $\mu\gtrsim 1.3$, suggesting the existence of a tricritical point (TCP) at $\mu\simeq 1.3$. The SF to FO transition is of second order type in all the frustrated region $(K>0)$ while it is first order for the unfrustrated regime till $K\simeq -2$, where a potential TCP exists.

In Fig. $\ref{CLcuts}$ we show the energy, condensate density, total density and bond-chiral OP for the four cuts ($a$, $b$, $c$ and $d$) displayed in the phase diagram of Fig. $\ref{CLdiag}$. Panels are labeled according to the corresponding cuts. Panel (a) shows the continuous vanishing of the bond-chiral OP along the half filling line, signaling a second order phase transition at $K_c=2$. Panel (b) shows the total density and the condensate density across the SF to FO transition along $K=2$. The condensate density vanishes continuously at $\mu_c=4$ characterizing a second order phase transition. For $\mu\gtrsim 1.3$, the SF to CSF transition is of first order type. Panel (c) displays the condensate and total densities along the $\mu=2$ line. Both quantities present a discontinuity at $K=5.6$, signaling a first order phase transitions. Panel (d) displays the crossing of the FO and CSF energies determining a first order phase transition.

In the next section we study the quantum phase diagram by means of the HMFT.

\section{\label{HMFT}Hierarchical mean-field theory}
%
HMFT \cite{hmft,isaevJQ} offers a simple but insightful scheme in which the inclusion of quantum correlations is carried out by identifying the relevant degrees of freedom in order to describe the different phases of interest. These new degrees of freedom define the hierarchical language \cite{adinphy} appropriate to describe the emergent phenomena. For the present case of the $J$-$K$ model on the square lattice, we will implement HMFT by tiling the lattice with clusters of equal size and shape $(L\times L)$. Being these clusters the new degrees of freedom, we can represent the many-body states of the cluster Fock space by CB creation  operators \cite{cbmft} over a vacuum of a new enlarged Fock space, in a similar way as it is done in other slave-particle approaches \cite{kr,aa,za}. The physical subspace of this enlarged CB Fock space is defined by all those many-body CB states which have one and only one CB on each site of the superlattice. Therefore, it is necessary to implement a physical constraint in order to obtain a physically meaningful solution. Within the physical subspace, the mapped Hamiltonian is exact. However, the new Hamiltonian is equally hard to solve as the original one and thus suitable many-body approximations are required. The advantage resides on the automatic inclusion of exact short-range quantum correlations when expressing the Hamiltonian in terms of the new CB operators. The physical constraint is usually treated in an approximated manner by the standard techniques, leading to states which admix physical and unphysical subspaces. Nevertheless, in this work we are going to restrict ourselves to the lowest order HMFT approximation, that is, a cluster Gutzwiller approximation which preserves the physical constraint exactly and therefore does not suffer from this inconvenience.

\begin{figure}[b]
\includegraphics[angle=0,clip=true,width=\columnwidth]{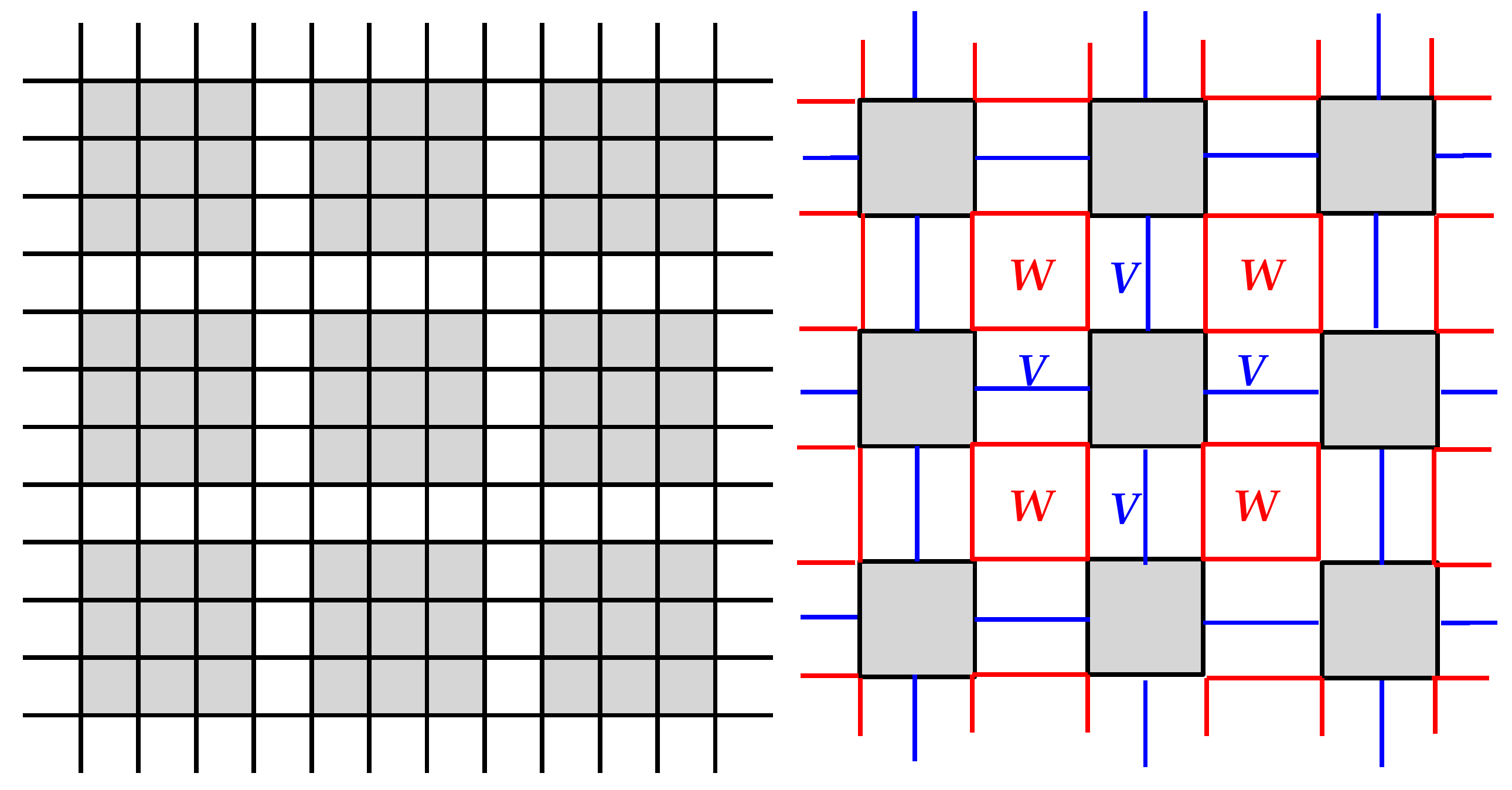}
\caption{\label{HMFT44scheme}(Color online)
Sketch of the 4$\times$4 tiling (left) and the resulting cluster superlattice (right) in which two-body (blue) and four-body interactions (red) are schematically represented by links connecting the clusters (shaded squares) and labeled by their corresponding matrix elements, $V$ and $W$ (see Eq.$(\ref{ccbh})$ in the text).}
\end{figure}

Let us start by mapping the original bosonic hard-core operators $\{a^{\dag}_{j},a_{j}\}$ to the new set of CBs \cite{cbmft,hmft},
\begin{equation}
a^{(\dag)}_{j}= \sum_{\mathbf{n}\mathbf{m}}\langle R\mathbf{n} \vert  a^{(\dag)}_{j} \vert R\mathbf{m} \rangle~ 
b^{\dag}_{R\mathbf{n}}b_{R\mathbf{m}},~~j\in R,
\label{cmap}
\end{equation}
where $\mathbf{n}\equiv (n_1,\ldots,n_{L^{2}})$ labels the occupation configuration of each cluster at superlattice site $R$. The new set of CB operators $\{b^{\dag}_{R\mathbf{n}},b_{R\mathbf{n}}\}$ obey the bosonic canonical commutation relations, and must satisfy the above mentioned physical constraint at each superlattice site, $\sum_{\mathbf{n}}b^{\dag}_{R\mathbf{n}}b_{R\mathbf{n}}=1$. As a consequence of the canonical mapping $(\ref{cmap})$, any operator $\hat{O}_{R}$ which is an algebraic function of the original hard-core bosonic operators $\{a^{\dag}_{i},a_{i}\}$ acting on sites which lie within a single cluster at position $R$ $(i,j,\ldots\in R)$ maps onto a one-body CB operator, 
\begin{equation}
\hat{O}_{R}=\sum_{\mathbf{m,n}}\langle R\mathbf{n}\vert \hat{O}_{R}\vert R\mathbf{m}\rangle~ b^{\dag}_{R\mathbf{n}}b_{R\mathbf{m}}. 
\end{equation}
Moreover, any operator which acts on $n$ different clusters of the superlattice $(i_1,j_1,\ldots\in R_1;~ i_2,j_2,\ldots\in R_2;~ i_n,j_n,\ldots\in R_n)$ maps onto an $n$-body CB operator, 
\begin{eqnarray}
\hat{O}_{R}&=&\sum_{\{\mathbf{m,n}\}}\langle R_1\mathbf{n}_1;\ldots;R_n\mathbf{n}_n\vert \hat{O}_{R}\vert R_1\mathbf{m}_1;\ldots;R_n\mathbf{m}_n\rangle \notag\\
&&\times b^{\dag}_{R_1\mathbf{n}_1}\ldots b^{\dag}_{R_n\mathbf{n}_n} b_{R_1\mathbf{m}_1}\ldots b_{R_n\mathbf{m}_n}. 
\end{eqnarray}
Applying this procedure to Hamiltonian $(\ref{JK})$, we obtain the CB Hamiltonian,
\begin{widetext}
\begin{eqnarray}
H_{CB}&=& \sum_{R}\sum_{\mathbf{n,m}}\langle R\mathbf{n} \vert H^{\square}\vert R\mathbf{m} \rangle~ b^{\dag}_{R\mathbf{n}}b_{R\mathbf{m}}
+ \sum_{\langle R_{1}R_{2}\rangle}
\sum_{\{\mathbf{n,m}\}}
\langle R_1\mathbf{n}_1;R_2\mathbf{n}_2 \vert H^{\parallel}
\vert R_1\mathbf{m}_1;R_2\mathbf{m}_2\rangle~ 
b^{\dag}_{R_1\mathbf{n}_1}b^{\dag}_{R_2\mathbf{n}_2}b_{R_1\mathbf{m}_1}b_{R_2\mathbf{m}_2}\nonumber\\
&&+\sum_{\langle R_1R_2R_3R_4 \rangle}
\sum_{\{\mathbf{n},\mathbf{m}\}}
\langle R_{1}\mathbf{n}_1;R_{2}\mathbf{n}_2;R_{3}\mathbf{n}_3;R_{4}\mathbf{n}_4 \vert 
H^{\times}
\vert R_{1}\mathbf{m}_1;R_{2}\mathbf{m}_2;R_{3}\mathbf{m}_3;R_{4}\mathbf{m}_4 \rangle \nonumber\\
&&~~\times b^{\dag}_{R_1\mathbf{n}_1}b^{\dag}_{R_2\mathbf{n}_2}b^{\dag}_{R_3\mathbf{n}_3}b^{\dag}_{R_4\mathbf{n}_4}b_{R_1\mathbf{m}_1}b_{R_2\mathbf{m}_2}b_{R_3\mathbf{m}_3}b_{R_4\mathbf{m}_4},\label{H_cb}
\end{eqnarray}
\end{widetext}
where $H^{\square}$ refers to all terms of Hamiltonian $(\ref{JK})$ acting within a cluster at site $R$ of the superlattice, $H^{\parallel}$ refers to all hopping and ring-exchange terms acting on sites of the original lattice belonging to two neighboring clusters $\langle R_{1}R_{2}\rangle$, and $H^{\times}$ denotes all ring-exchange terms which act on sites belonging to four neighboring clusters $\langle R_{1}R_{2}R_{3}R_{4}\rangle$. Let us now apply a general unitary transformation among the $b^{(\dag)}_{R\mathbf{n}}$ bosons, i.e., $b^{(\dag)}_{R\mathbf{n}}=\sum_{\alpha}U^{\alpha(\ast)}_{R\mathbf{n}}b^{(\dag)}_{R\alpha}$, where the greek indices label a new orthonormal basis. We then arrive to a general CB Hamiltonian of the form
\begin{eqnarray}
H_{CB}&=& \sum_{R} \left(T_{R}\right)^{\alpha}_{\beta} b^{\dagger}_{R\alpha}b_{R\beta}\label{ccbh}\\
&&+ \sum_{\langle R_1 R_2\rangle}\left(V_{R_1R_2}\right)^{\alpha_1\alpha_2}_{\beta_1 \beta_2}
b^{\dagger}_{R_1\alpha_1}b^{\dagger}_{R_2\alpha_2}b_{R_1\beta_1}b_{R_2\beta_2}\nonumber\\
&& +\sum_{\langle R_1 R_2 R_3 R_4\rangle}
\left(W_{R_1R_2R_3R_4}\right)^{\alpha_1\alpha_2\alpha_3\alpha_4}_{\beta_1\beta_2\beta_3\beta_4}\notag\\
&&~\times 
b^{\dagger}_{R_1\alpha_1}b^{\dagger}_{R_2\alpha_2}b^{\dagger}_{R_3\alpha_3}b^{\dagger}_{R_4\alpha_4}
b_{R_1\beta_1}b_{R_2\beta_2}b_{R_3\beta_3}b_{R_4\beta_4},\notag
\end{eqnarray}
where repeated greek indices are summed over. All the information about the original Hamiltonian, the tiling of the lattice and the rotation $\hat{U}$, is contained in the tensors $\hat{T}$, $\hat{V}$ and $\hat{W}$, as it is schematically represented in Fig. \ref{HMFT44scheme}. It is straightforward, though lengthy, to explicitly write the form of the tensors $\hat{T}$, $\hat{V}$, and $\hat{W}$; we refer the reader to Appendix \ref{matrixel}.

%
Hamiltonian (\ref{ccbh}) is an exact mapping of the original Hamiltonian (\ref{JK}), provided that the physical constraint is satisfied in each cluster. Nevertheless, it is equally hard to solve as the original one. It serves though as a good starting point for different approximations which will capture the short-range real-space correlations automatically. We next apply the Gutzwiller approximation which preserves the physical constraint exactly. 
\subsection{\label{gutz}Gutzwiller approximation}
%
Within Gutzwiller's approximation, the variational ansatz is a cluster product wave function,
\begin{equation}
\vert \Psi_{0} \rangle =\prod_{R}\vert \Phi_R\rangle=\prod_{ R } \left(\sum_{\mathbf{n}}U_{R\mathbf{n}}^{\sf{g}} 
b_{R\mathbf{n}}^{\dag} \right)\vert 0\rangle,
\label{HMFTwf}
\end{equation}
where the label $\sf{g}$ stands for \textit{ground state}. The components of $U_{R\mathbf{n}}^{\sf{g}}$ will be determined upon minimization of the energy. As the ansatz (\ref{HMFTwf}) preserves the physical constraint exactly, the energy obtained will be an upper bound to the exact one, or in other words, variational \cite{hmft}. Assuming a uniform ground state on the superlattice, i.e., $\hat{U}_{R}=\hat{U}$, the energy per site computed with the Gutzwiller wave function $(\ref{HMFTwf})$ is 
\begin{equation}
E=\left(T^{\sf{g}}_{\sf{g}} + 2V^{\sf{gg}}_{\sf{gg}} + W^{\sf{gggg}}_{\sf{gggg}}\right)/L^{2}.\label{E_hmf}
\end{equation}
Adding a Lagrange multiplier $\lambda$ to preserve the normalization of our variational wave function (\ref{HMFTwf}), $\sum_{\mathbf{n}} U^{\sf{g}\ast}_{\mathbf{n}}U^{\sf{g}}_{\mathbf{n}}=1$, we proceed to minimize the energy, 
\begin{eqnarray}
\frac{\partial}{\partial U^{\sf{g}\ast}_{\mathbf{m}}}
\left( E[\{U^{\sf{g}},U^{\sf{g}\ast}\}] - \lambda\sum_{\mathbf{n}}U^{\sf{g}\ast}_{\mathbf{n}}U^{\sf{g}}_{\mathbf{n}} \right)=0.\label{variational}
\end{eqnarray}
The resulting equation can be rewritten as a Hartree eigensystem of the general form
\begin{equation}
\sum_{\mathbf{n}}h_{\mathbf{mn}} U^{\sf{g}}_{\mathbf{n}}=\lambda U^{\sf{g}}_{\mathbf{m}},\label{Diag_hartree}
\end{equation}
where $\hat{h}$ is the Hartree matrix and $\lambda$ and $U^{\sf{g}}$ are the lowest eigenvalue and corresponding eigenvector, respectively. The rest of the eigenvectors obtained in the diagonalization procedure define a basis in the orthogonal space to the ground state ($\sf{g}$).

Due to the algebraic dependence of the Hartree matrix on the amplitudes $U^{\sf{g}}_\mathbf{n}$, Eq.  (\ref{Diag_hartree}) comprises  a set of nonlinear equations which is solved iteratively after starting with an initial guess for the amplitudes. Being a variational procedure, the energy decreases at each iteration step, converging to a minimum at selfconsistency. Notice that the use of finite clusters, chosen to be $L\times L$ squares in the 
present case, allows for the description of a wide range of multiple-sublattice phases.

\begin{figure}[t]
\includegraphics[angle=0,clip=true,width=\columnwidth]{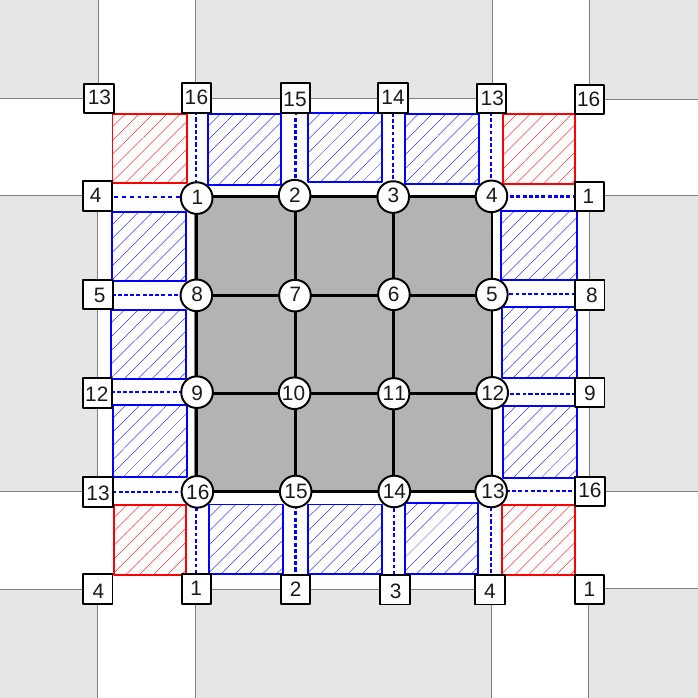}
\caption{(Color online) Sketch showing a 4$\times$4 cluster with OBC embedded in its mean-field environment. Numbers label the sites within the cluster (circles) and its vicinity (squares). Exact hopping and ring-exchange interactions within the cluster are represented by solid lines and dark shade. The chemical potential term acts on sites within the cluster. The auxiliary  fields $(\psi_{i}^{\ast},\psi_{i},\varphi_{ij}^{\ast},\varphi_{ij})$, which account for the mean-field embedding (see text), are evaluated on the squared sites belonging to the boundaries of the nearby clusters. The mean-field ring-exchange among four (two) neighbouring clusters is represented by red (blue) grid plaquettes, and symbolized within the formalism by $h^{\times}$ ($h^{\parallel}$). The mean-field hopping interaction is represented by dotted lines.}
\label{hartree_schem}
\end{figure}

Solving the Hartree eigensystem (\ref{Diag_hartree}) is equivalent to the exact diagonalization of a cluster of size $L$$\times$$L$ with OBC and a set of self-consistently defined auxiliary fields acting on its boundaries, which mimic the environment in the mean-field approximation. We can therefore express the Hartree matrix $\hat{h}$ as a sum of intra- and inter-cluster
terms,
\begin{eqnarray}
\hat{h}&=& \hat{h}^{\square} + \hat{h}^{\parallel} + \hat{h}^{\times}\label{hmf}.
\end{eqnarray}
The intra-cluster terms are all hopping, ring-exchange and chemical potential terms which act within the $L$$\times$$L$ cluster (all parameters in units of $J$),
\begin{eqnarray}
\hat{h}^{\square}&=& 
-\sum_{\langle ij\rangle\in\square}(a^{\dag}_{i}a_{j} + h.c.)  
-\mu\sum_{j\in\square}n_{j}\notag\\
&&+ K\sum_{\langle ijkl\rangle\in\square}(a^{\dag}_{i}a^{\dag}_{k}a_{j}a_{l} + h.c.)\label{hsquare}.
\end{eqnarray}
The mean-field interaction among two nearest neighbour clusters leads to
\begin{eqnarray}
\hat{h}^{\parallel}&=&-\sum_{\langle ij\rangle\in\parallel}(a^{\dag}_{i}\psi_{j} + a_{i}\psi_{j}^{\ast})\notag\\
&&+K\sum_{\langle ijkl\rangle\in\parallel}(a^{\dag}_{i}a_{j}\varphi_{kl}^{\ast}+a^{\dag}_{j}a_{i}\varphi_{kl}),\label{hparallel}
\end{eqnarray}
where the sums are restricted to bonds, in the first case, and plaquettes, in the second (see Fig. \ref{hartree_schem}). That is, creation (annihilation) hard-core bosonic operators act on sites lying on the boundaries of the cluster while the auxiliary fields are evaluated on the boundaries of the neighbouring cluster. In the same way, the ring-exchange interaction among four clusters leads to
\begin{eqnarray}
\hat{h}^{\times}&=&
K\sum_{\langle ijkl\rangle\in\times}(a^{\dag}_{i}\psi_{j}\psi_{k}^{\ast}\psi_{l} + a_{i}\psi_{j}^{\ast}\psi_{k}\psi_{l}^{\ast}),\label{htimes}
\end{eqnarray}
where now the sum reduces to the four plaquettes which touch the four corners of the cluster (see Fig. \ref{hartree_schem}). Bosonic creation (annihilation) operators act on the four corners of the cluster and are coupled to three external auxiliary fields evaluated at the corners of the corresponding neighbouring clusters. 
%

The auxiliary fields are self-consistently defined by
\begin{eqnarray}
\psi_{j}^{\ast}&=& \langle \Phi\vert a^{\dag}_{j}\vert \Phi\rangle
=\sum_{\mathbf{n'}}~ U^{\sf{g}\ast}_{\lbrace 1_{j}\rbrace} U^{\sf{g}}_{\lbrace 0_{j}\rbrace},\label{psi}\\
\varphi_{ij}^{\ast}&=&\langle \Phi\vert a^{\dag}_{i}a_{j}\vert \Phi\rangle=
\sum_{\mathbf{n'}}~ U^{\sf{g}\ast}_{\lbrace1_{i}0_{j}\rbrace} U^{\sf{g}}_{\lbrace 0_{i}1_{j}\rbrace}\label{varphi},
\end{eqnarray}
where $\vert \Phi\rangle$ is the cluster wave function defined in Eq. $(\ref{HMFTwf})$, and  $\{1_i,0_j\}\equiv (n_1,\ldots,1_i,0_j,\ldots,n_{L^2})$ refers to a cluster configuration $\mathbf{n}$ with the occupation of sites $i$ and $j$ fixed to be $1$ and $0$, respectively. The sums in (\ref{psi}) and (\ref{varphi}) run over the configurations of all sites within the cluster except those at which the field is evaluated.

The energy per site $(\ref{E_hmf})$ in units of $J$ can be equivalently written in terms of the lowest eigenvalue $\lambda$ of the Hartree eigensystem $(\ref{Diag_hartree})$ and the auxiliary fields $\left\{\psi,\varphi\right\}$ as
\begin{eqnarray}
E&=& \frac{1}{L^2}[\lambda+
\frac{1}{2}\sum_{\langle ij\rangle\in\parallel}\left( \psi^{\ast}_{i}\psi_j + \psi^{\ast}_{j}\psi_i \right)\notag\\ 
&&- \frac{K}{2}\sum_{\langle ijkl\rangle\in\parallel}\left( \varphi^{\ast}_{ij}\varphi_{kl}+\varphi_{kl}^{\ast}\varphi_{ij} \right)\notag\\
&&- \frac{3K}{4}\sum_{\langle ijkl\rangle\in\times}\left( \psi^{\ast}_{i}\psi^{\ast}_{k}\psi_{j}\psi_{l} + \psi^{\ast}_{j}\psi^{\ast}_{l}\psi_{i}\psi_{k}\right)],
\end{eqnarray}
where we subtract to the Hartree eigenvalue $\lambda$ double counting terms coming from the two- and four-cluster interactions.

In the limit $L=1$, the superlattice and the original lattice are exactly the same and this approach is equivalent to the classical approximation derived in Sec.\ref{ClasPhaseDiag}, account taken of the two-sublattice structure, i.e., with $U_j^{\ast}= U_i$ for $i\in A,~j\in B$. In this limit, the mapping (\ref{cmap}) applied to hard-core bosons reduces to the Schwinger boson mapping of $SU(2)$ spin operators \cite{auerbach} written in the bosonic language. As we have seen, the matrix $\hat{U}$ automatically splits the ground state flavor $(\sf{g})$ from its orthogonal space at each superlattice site. Within linear spin-wave theory (LSWT), the relevant quantum fluctuations over a semiclassical ground state  are assumed to reside in its orthogonal space. Thus, HMFT offers a convenient scheme for computing low-lying excitations over multiple-sublattice classical ground states of Hamiltonians with highly non-trivial interaction terms, as it is the case for the CSF phase present in our ring-exchange model. In Appendix \ref{LSWT} we provide details of the computation of LSWT excitations of the classical phase diagram derived in Sec.\ref{ClasPhaseDiag} by means of this method.

\subsection{\label{OPs}Order parameters and observables}
%
In order to characterize the phases we compute within HMFT ($L=2,4$) the $(\pi,\pi)$ CDW order parameter and the two OPs already defined in the previous section, i.e., the condensate density at $\mathbf{k}=(0,0)$ (\ref{cond_dens}) and the bond-chiral OP (\ref{bchop}). We also compute the expectation values of the hopping (\ref{B}) and plaquette (\ref{P}) operators over the lattice to characterize the various solid phases obtained. 

The condensate density computed with the Gutzwiller wave function (\ref{HMFTwf}) in the thermodynamic limit is
\begin{equation}
\rho_{0}= \frac{1}{N^{2}}
\left(\sum_{R}\sum_{i\neq j}\langle a_{i}^{\dag}a_{j}\rangle + \sum_{R\neq R'}\sum_{i\neq j}
\langle a_{i}^{\dag}\rangle\langle a_{j}\rangle\right),\label{HMFrho}
\end{equation}
where $i,j$ lie within the same cluster $R$ in the first term, and $i\in R$ and $j\in R'\neq R$ in the second. The first term vanishes in the thermodynamic limit for clusters of finite size, leading to
\begin{eqnarray}
\rho_{0}=\frac{1}{L^{4}}\sum_{i\in \square}\langle a_{i}^{\dag}\rangle \sum_{j\in \square}\langle a_{j}\rangle=\frac{1}{L^4}\vert \sum_{i\in\square} \psi_{i}^{\ast}\vert^{2},\label{rho_HMF}
\end{eqnarray}
where we took into account that the number of clusters is $M=N/L^{2}$ and used the definition of the auxiliary field $\psi^{\ast}_{j}$ in (\ref{psi}). 

\begin{figure*}[t]
\includegraphics[width=2\columnwidth]{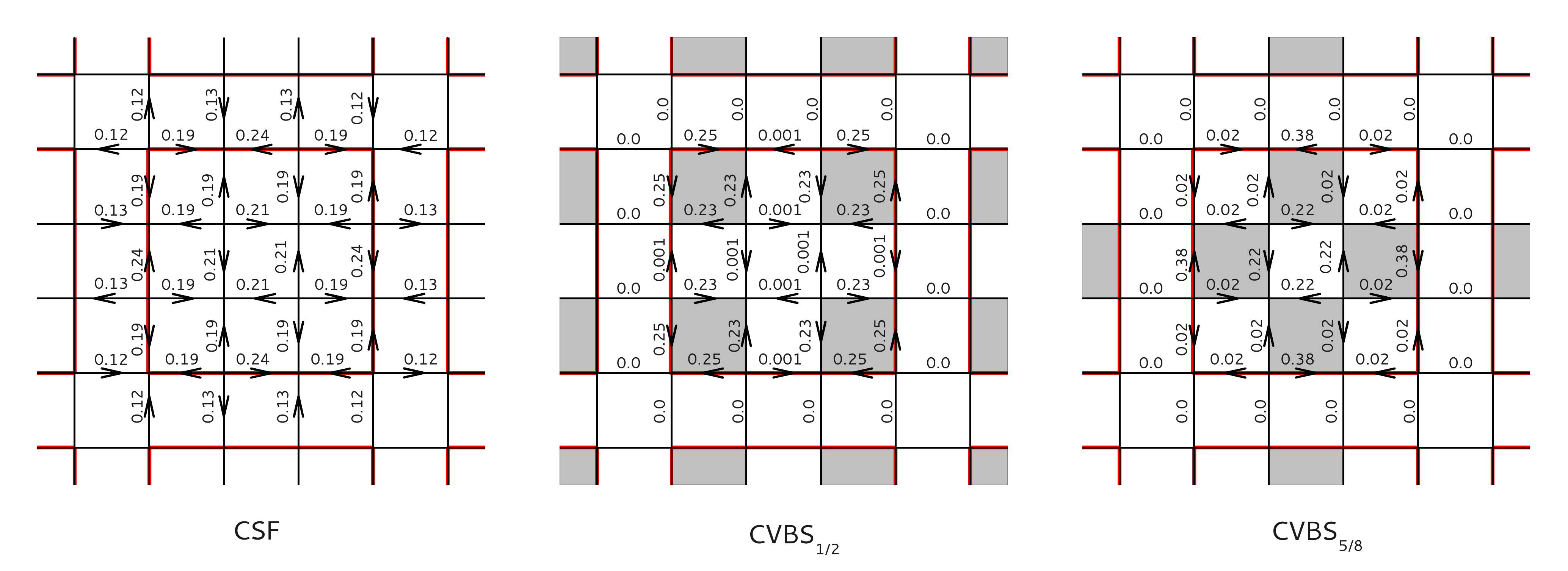} 
\caption{(Color online) Schematic picture showing the CSF, CVBS$_{1/2}$ and CVBS$_{5/8}$ bond-chiral phases (from left to right). The arrows indicate the bond-currents and the numbers the magnitude of the bond-chiral expectation value for each intra- and inter-cluster bond computed with HMFT--4$\times$4 at $\mu=0,~K=2.3$ (CSF), $\mu=0,~K=10$ (CVBS$_{1/2}$) and $\mu=3.5,~K=5.6$ (CVBS$_{5/8}$). Grey squares highlight the underlying plaquette structure of the two CVBS phases. The magnitude of the bond-chiral order is almost uniform in the CSF phase while it has an alternating plaquette pattern in the solid phases.}
\label{chiral}
\end{figure*}
The bond-chiral OP computed within the Gutzwiller approximation leads to a sum of intra- and inter-cluster terms
\begin{equation}
\Omega= \frac{1}{N_{b}}\left(\sum_{\langle ij\rangle} \vert\hat{\Omega}_{ij}^{\square} \vert
+\sum_{\langle ij\rangle} \vert\hat{\Omega}_{ij}^{\parallel} \vert\right).
\end{equation}
where $N_{b}=2N$ is the total number of bonds. The first sum runs over all bonds lying within the clusters and the second one over all bonds linking two different clusters. The expectation value of the bond-chiral operator (\ref{bo_clas}) acting on a bond $\langle ij\rangle$ lying within a cluster is
\begin{equation}
\langle\hat{\Omega}_{ij}^{\square}\rangle=-\Im(\varphi^{\ast}_{ij}),\label{omsquare}
\end{equation}
where $\Im(z)$ refers to the imaginary part of a complex scalar $z$ and $\varphi^{\ast}_{ij}$ is the auxiliary field defined in (\ref{varphi}). The expectation value of the bond-chiral operator (\ref{bo_clas}) acting on a bond $\langle ij\rangle$ which links two neighbouring clusters is  
\begin{equation}
\langle\hat{\Omega}_{ij}^{\parallel}\rangle=-\Im(\psi_{i}^{\ast}\psi_{j}) , \label{omparal}
\end{equation}
where $\psi_{i}^{\ast}$ is the auxiliary field defined in (\ref{psi}).

The $(\pi,\pi)$ CDW order parameter is defined as the normalized spin structure factor at wave vector $\mathbf{q}=(\pi,\pi)$, 
\begin{equation}
M_{s}^{2} = S(\pi,\pi)/N,\label{ms}
\end{equation}
where the spin structure factor is defined as the two-point correlator of $S^{z}$ at equal momentum, i.e., $S(\mathbf{q})=\sum_{ij} e^{\mathrm{i}(\mathbf{r}_i-\mathbf{r}_j)\mathbf{q}}\langle S^{z}_i S^{z}_j \rangle/N$. Following similar arguments as we did for the computation of the condensate density (\ref{HMFrho}), Eq. (\ref{ms}) simplifies, in the thermodynamic limit, to
\begin{equation}
M_{s}=\frac{1}{L^2}\sum_{j\in\square}e^{\mathrm{i} (\pi,\pi) \mathbf{r}_j}\langle n_j - 1/2 \rangle , \label{mstag}
\end{equation}
where $\mathbf{r}_i$ is the position of site $i$ within the cluster and we have rewritten $S^{z}$ in the bosonic language.

Equivalently, the expectation values of the hopping and plaquette operators depend on whether they act on sites inside a cluster or connecting different clusters. Thus, for the hopping operator, we are led to the expressions
\begin{eqnarray}
\langle \hat{B}_{ij}^{\square}\rangle&=& 2\Re(\varphi_{ij}^{\ast})
\label{Bsquare}
\end{eqnarray}
and
\begin{eqnarray}
\langle \hat{B}_{ij}^{\parallel}\rangle&=&2\Re(\psi_{i}^{\ast}\psi_{j}),
\label{Bparal}
\end{eqnarray}
depending on whether the bond $\langle ij\rangle$ is inside a cluster or is shared by two clusters, respectively. We have labeled with $\Re(z)$ the real part of a complex scalar $z$ and we have made use of the auxiliary fields $\psi^{\ast}$ and $\varphi^{\ast}$ defined in (\ref{psi}) and (\ref{varphi}). Note that the expectation values of the hopping (\ref{B}) and bond-chiral (\ref{bo_clas}) operators are directly related to the real and imaginary parts of the expectation value of a single hopping process, i.e., $\langle a^{\dag}_{i}a_{j}\rangle$. 

Finally, the expectation values of the plaquette operator are
\begin{eqnarray}
\langle P_{ijkl}^{\square}\rangle&=&
2\Re\left( \sum_{\mathbf{n'}}~U^{\sf{g}\ast}_{\{1_i,0_j,1_k,0_l\}}U^{\sf{g}}_{\{0_i,1_j,0_k,1_l\}} \right),\label{Psquare}\\
\langle P_{ijkl}^{\parallel}\rangle&=&
2\Re(\varphi^{\ast}_{ij}\varphi_{kl}),\label{Pparal}
\end{eqnarray}
or
\begin{eqnarray}
\langle P_{ijkl}^{\times}\rangle&=&
2\Re(\psi^{\ast}_i\psi_j\psi^{\ast}_k\psi_l),\label{Pcross}
\end{eqnarray}
depending on whether $\hat{P}_{ijkl}$ acts on a plaquette lying within the cluster (\ref{Psquare}), between two clusters (\ref{Pparal}) or connecting four clusters (\ref{Pcross}). In the first case, the sum is restricted to the configurations over all sites within the cluster except those belonging to the plaquette $\langle ijkl\rangle$ at which the operator $\hat{P}_{ijkl}$ is evaluated. 

\subsection{\label{HMFphasediag}Description of different valence bond phases}
%
Using clusters of size 2$\times$2 and 4$\times$4 as the new degrees of freedom allows us to access several plaquette  phases which cannot be described by standard mean-field techniques. Apart from the three phases already obtained by means of the classical approximation (FO, SF and CSF), HMFT unveils three more phases: a valence bond solid phase for $K<0$, and two novel valence bond-chiral solid phases for $K>0$.
\subsubsection{Valence bond solid $\rho=1/2$ (VBS):}
This phase is characterized by the alternating expectation value of the hopping and plaquette operators (\ref{Bsquare})-(\ref{Pcross}) along the $x$ and $y$ directions, fixed total density $\rho=1/2$ and absence of bond-chiral, superfluid, or $(\pi,\pi)$ CDW orders. Within the 2$\times$2 approximation, the wave function obtained is a linear combination of just the $4!/2!2!=6$ possible half filled configurations,
\begin{eqnarray}
&&\vert\Phi^{2\times 2}\rangle=\alpha\left(\left\vert\begin{array}{c}
\includegraphics[clip=true,scale=0.35]{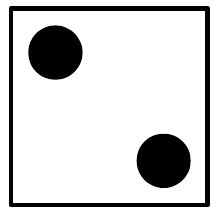}\end{array}
\right\rangle + 
\left\vert\begin{array}{c}
\includegraphics[clip=true,scale=0.35]{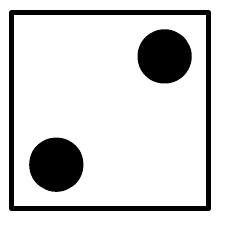}\end{array}
\right\rangle\right)\label{VBS1}\\
&&~+\beta\left(  
\left\vert\begin{array}{c}
\includegraphics[clip=true,scale=0.35]{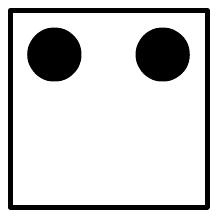}\end{array}
\right\rangle
+
\left\vert\begin{array}{c}
\includegraphics[clip=true,scale=0.35]{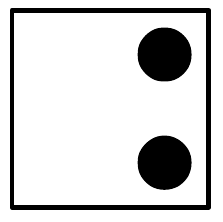}\end{array}
\right\rangle
+
\left\vert\begin{array}{c}
\includegraphics[clip=true,scale=0.35]{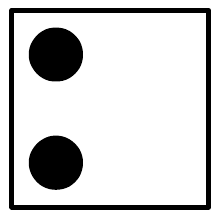}\end{array}
\right\rangle
+
\left\vert\begin{array}{c}
\includegraphics[clip=true,scale=0.35]{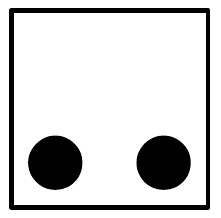}\end{array}
\right\rangle
\right)\notag,
\end{eqnarray}
where the amplitudes $\alpha$ and $\beta$ are real. In the spin language, this phase is paramagnetic, i.e., $\langle\mathbf{S}_{j}\rangle=0$. It preserves the global $U(1)$ and $C_{4}$ symmetries of the Hamiltonian (\ref{JK}). However, it mixes the total number of bosons in each row and column, breaking the row/column ($d=1$) $U(1)$ gauge-like symmetries (\ref{colrowsym}).
%
\begin{figure*}[t!]
\includegraphics[clip=true, width=1.9\columnwidth]{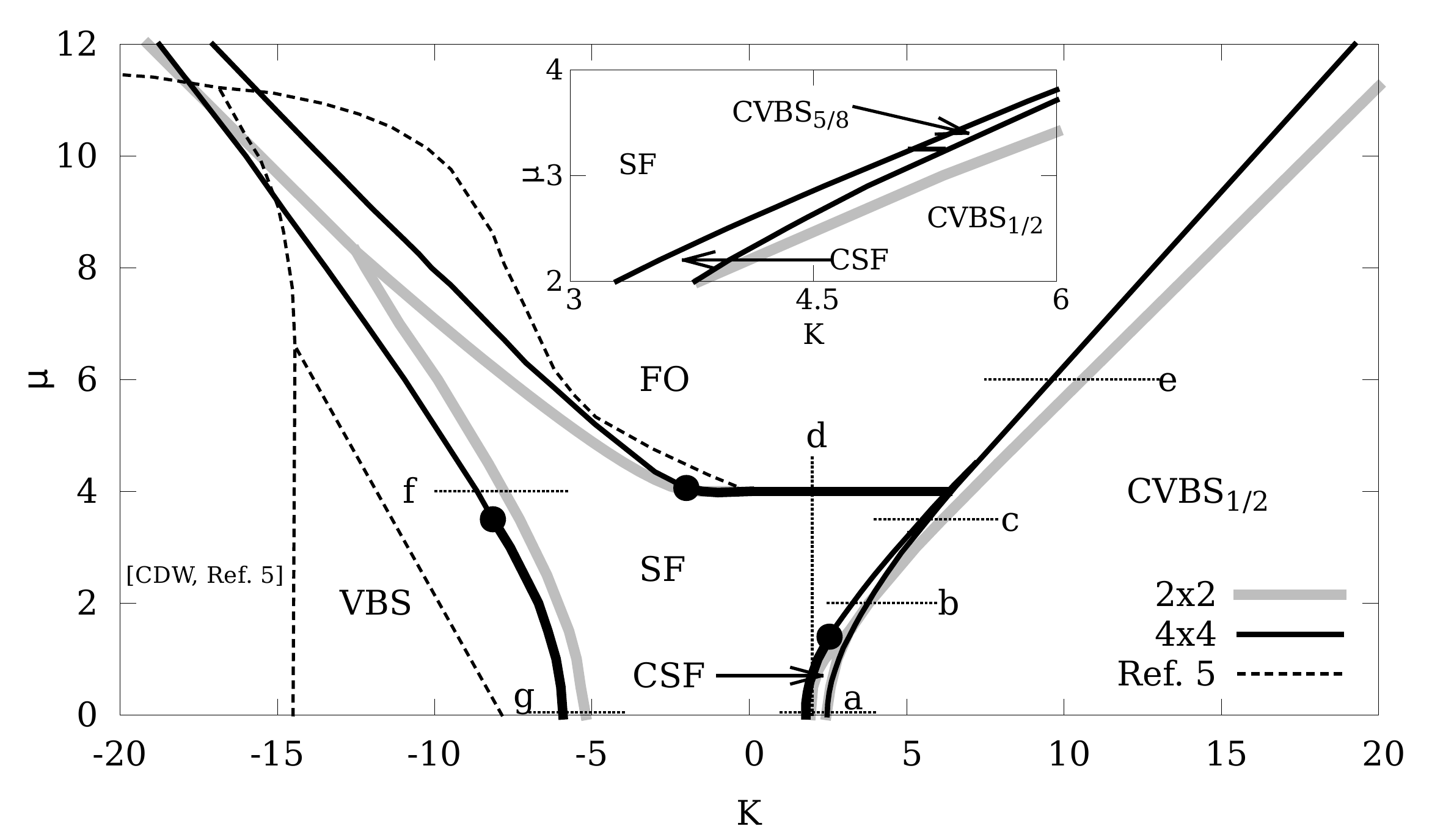}
\caption{HMFT--$L$$\times$$L$ phase diagram for $L=2$ (gray) and $L=4$ (black) together with QMC results from Ref. \onlinecite{qmc} (dashed line). Thick (thin) lines correspond to second (first) order phase transitions. Short dashed lines correspond to several cuts (lower case letters) for which we have examined the transition across the various phases obtained with HMFT (in capital letters). Phases from Ref. \onlinecite{qmc} are labelled in italics. The filled (black) circles correspond to potential tricritical points (TCP) within HMFT--4$\times$4. The corresponding TCPs within the HMFT--2$\times$2 are indistinguishable in the SF-FO and SF-CSF transitions, while the SF-VBS transition is always first order. We cannot discard the possibility that some of the second order transitions are weakly first order (see text).  Inset: zooming of the small region where the CVBS$_{5/8}$ phase emerges.}
\label{HMFTdiag}
\end{figure*}
\subsubsection{Half filled valence bond-chiral solid (CVBS$_{1/2}$):}
This phase is a bond-chiral counterpart of the VBS previously described. It preserves the $U(1)$ symmetry of Hamiltonian (\ref{JK}) but breaks $C_4$ down to $C_2$, as it can also be inferred by its {\it source-and-drain} chiral pattern. Apart from alternating expectation values of the hopping and plaquette operators (\ref{Bsquare})-(\ref{Pcross}) and null superfluid and $(\pi,\pi)$ CDW orders, it has nonzero bond-chiral order. The expectation value of the bond-chiral operator has a source-and-drain current pattern reminiscent of the CSF, as it is schematically represented in Fig. \ref{chiral}. In the spin language, this phase is a paramagnet, i.e., $\langle \mathbf{S}_j \rangle=0$, with nonzero spin chirality. The cluster wave function obtained within HMFT--2$\times$2 is equivalent to the previous VBS (\ref{VBS1}), but with complex amplitudes in the diagonal configurations,
\begin{eqnarray}
&&\vert\Phi^{2\times 2}\rangle=\alpha\left(e^{\mathrm{i}\varphi}\left\vert\begin{array}{c}
\includegraphics[clip=true,scale=0.35]{5.pdf}\end{array}
\right\rangle + 
e^{-\mathrm{i}\varphi}\left\vert\begin{array}{c}
\includegraphics[clip=true,scale=0.35]{10.pdf}\end{array}
\right\rangle\right)\label{CVBS2}\\
&&~+\beta\left(  
\left\vert\begin{array}{c}
\includegraphics[clip=true,scale=0.35]{3.pdf}\end{array}
\right\rangle
+
\left\vert\begin{array}{c}
\includegraphics[clip=true,scale=0.35]{6.pdf}\end{array}
\right\rangle
+
\left\vert\begin{array}{c}
\includegraphics[clip=true,scale=0.35]{9.pdf}\end{array}
\right\rangle
+
\left\vert\begin{array}{c}
\includegraphics[clip=true,scale=0.35]{12.pdf}\end{array}
\right\rangle
\right),\notag
\end{eqnarray}
where $\alpha$ and $\beta$  are real and $0<\varphi<\pi/2$. Moreover, in the $K$-only limit (\ref{Konly}) both the VBS and CVBS$_{1/2}$ wave functions have the same amplitudes, $\alpha=\sqrt{3/8}$ and $\beta=1/4$, with a phase $\varphi=\pi/2$. This is consistent with the chiral symmetry of the $K$-only Hamiltonian (\ref{Konly}) described in Sec.\ref{intro}.

Within the HMFT--4$\times$4 approximation, the cluster wave function obtained for both the VBS and CVBS$_{1/2}$ phases live in the subspace of the $16!/8!8!=12870$ half filled 4$\times$4 cluster configurations. Similarly to HMFT--2$\times$2, the amplitudes $U^{\sf{g}}_{\mathbf{n}}$ are real (complex) for the VBS (CVBS$_{1/2}$) phase. Nevertheless, the 4$\times$4 wave function preserves the alternating plaquette pattern already found by means of HMFT--2$\times$2, indicating that it introduces minor quantitative corrections over the 2$\times$2 description. Moreover, in the $K$-only limit, the number of nonzero amplitudes $U^{\sf{g}}_{\mathbf{n}}$ of the HMFT--4$\times$4 wave function is 1534. The leading amplitudes correspond to occupation configurations $\mathbf{n}$ which can be written as a direct product of four 2$\times$2 diagonal configurations, i.e.,
\begin{eqnarray}
\vert\Phi^{4\times 4}\rangle&=&
\widetilde{\alpha}\left(\left\vert\begin{array}{c}
\includegraphics[clip=true,scale=0.3]{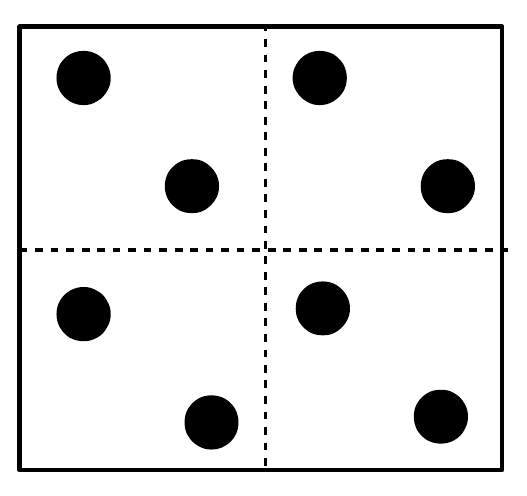}\end{array}
\right\rangle+
\left\vert\begin{array}{c}
\includegraphics[clip=true,scale=0.3]{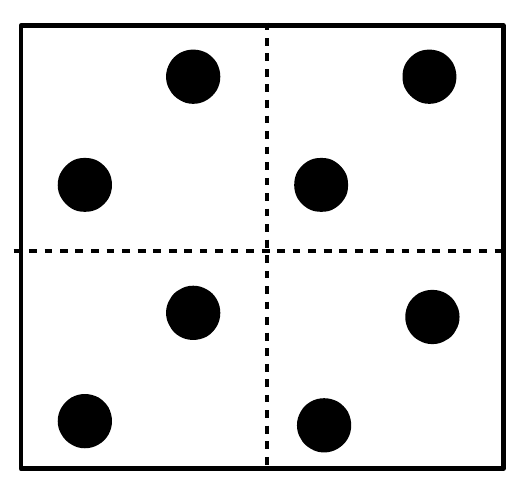}\end{array}
\right\rangle\right)\notag\\ 
&&+ 
\ldots,\label{cvbs12wf}
\end{eqnarray}
where $\ldots$ are the other relevant configurations. It is important to remark that at this $K$-only limit, the wave function obtained by exact diagonalization of a 4$\times$4 cluster with PBC has only 82 configurations (out of the original 12870) with nonzero amplitudes, all of them satisfying the $d=1$ gauge-like symmetries mentioned in Sec. \ref{intro}. 
\subsubsection{Valence bond-chiral solid $\rho=5/8$ (CVBS$_{5/8}$):}
%
HMFT--4$\times$4 results slightly modify those already found with HMFT--2$\times$2 with the exception of a small region of the phase diagram (see Fig. \ref{HMFTdiag}) where another valence bond-chiral solid phase with total density $\rho=5/8$ emerges. In the spin language, this is a magnetic phase, i.e., $\langle S^{z}_{j}\rangle=1/8$, with nonzero bond-chiral order. This particular solid phase cannot be captured within HMFT--2$\times$2 scheme as it has a density which is non-commensurate with the 2$\times$2 cluster size. The alternating plaquette pattern present in CVBS$_{1/2}$ changes (see Fig. \ref{chiral}) and the number of bonds with appreciable intensity of the expectation value of the bond-chiral operator diminshes. This is a manifest consequence of the doping, which allows for less hopping and ring-exchange processes over the system, as it can be deduced by inspecting the most relevant components of the resulting 4$\times$4 cluster wave function, 
\begin{eqnarray}
\vert\Phi^{4\times 4}\rangle&=&
\gamma\left(e^{\mathrm{i}\eta}\left\vert\begin{array}{c}
\includegraphics[clip=true,scale=0.3]{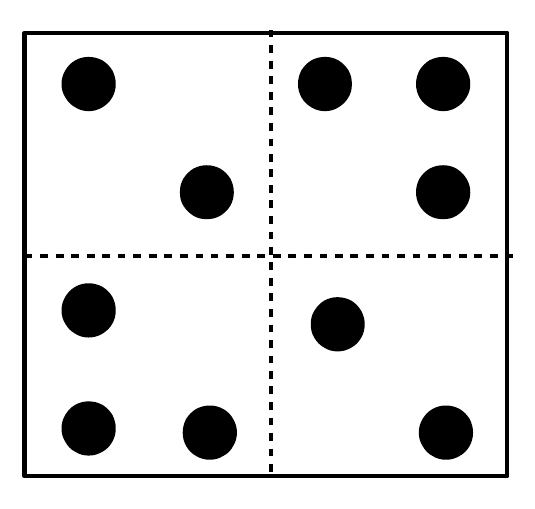}\end{array}
\right\rangle+
e^{-\mathrm{i}\eta}\left\vert\begin{array}{c}
\includegraphics[clip=true,scale=0.3]{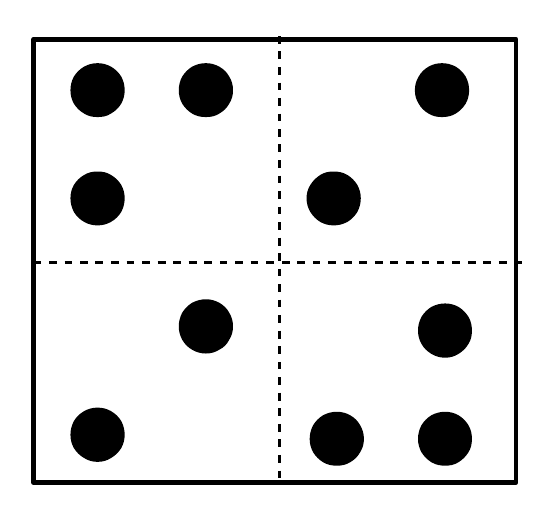}\end{array}
\right\rangle\right)\notag\\ 
&&+ 
\ldots\label{cvbs58wf}
\end{eqnarray}
where $\gamma$ is real and $0<\eta<\pi/2$. Notice that these configurations are related to the ones in (\ref{cvbs12wf}) by the addition of two bosons at the corners of the cluster, thus maximizing the number of available ring-exchange processes within the 4$\times$4 cluster while preserving the $C_2$ symmetry.
%
\begin{figure}[!]
\includegraphics[clip=true,width=\columnwidth]{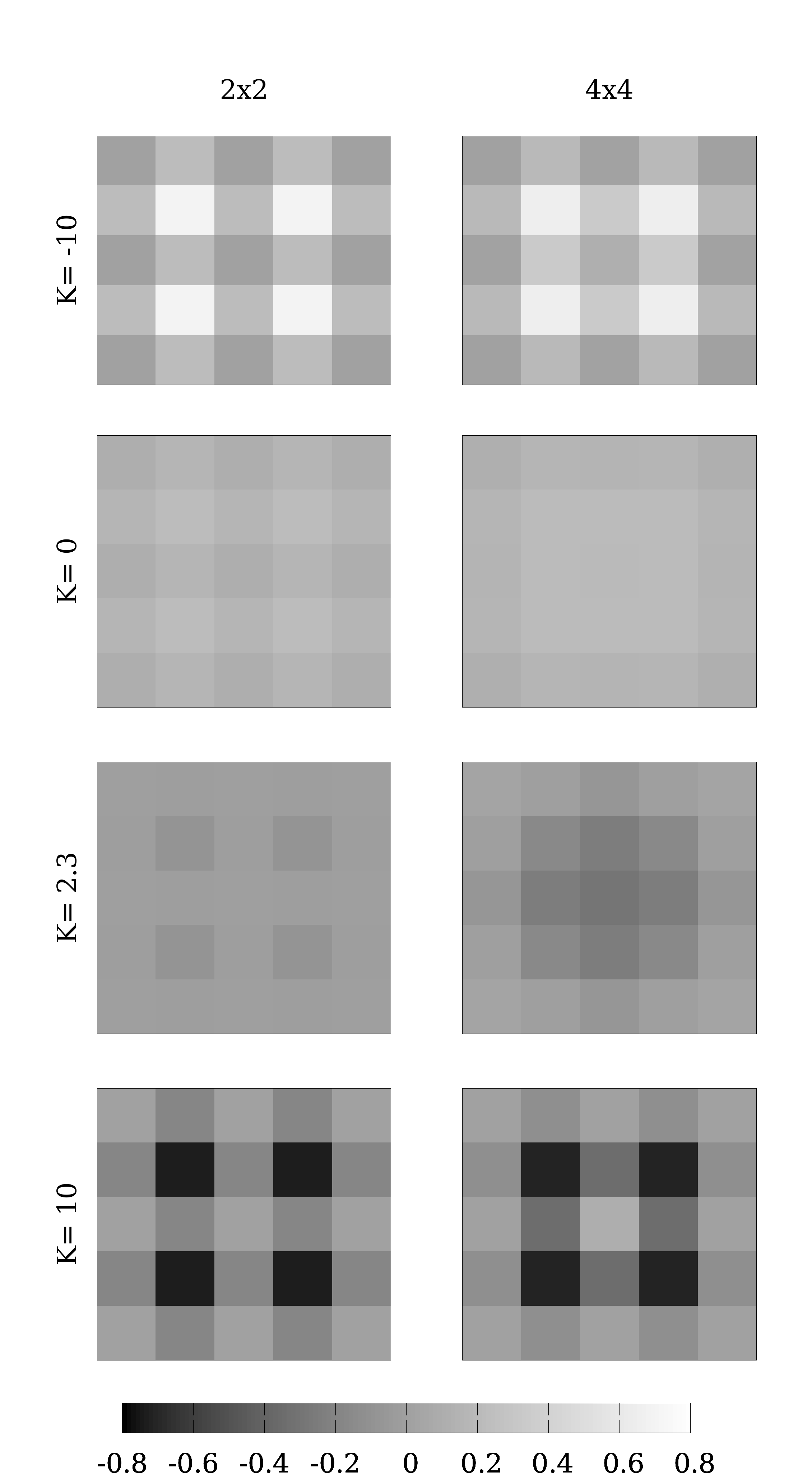}
\caption{\label{plaqOP}
Expectation value of the plaquette operator (\ref{P}) computed within HMFT--2$\times$2 (left) and HMFT--4$\times$4 (right) for several values of $K$ along the half filling line $(\mu=0)$: $K=-10$ (VBS), $K=0$ (SF), $K=2.3$ (CSF), $K=10$ (CVBS$_{1/2}$). The panels display four $2\times 2$ clusters surrounded by the inter-cluster plaquettes (left) and the corresponding $4\times 4$ cluster surrounded by the inter-cluster plaquettes (right).}
\end{figure}
%
\begin{figure}[!]
\includegraphics[clip=true,width=\columnwidth]{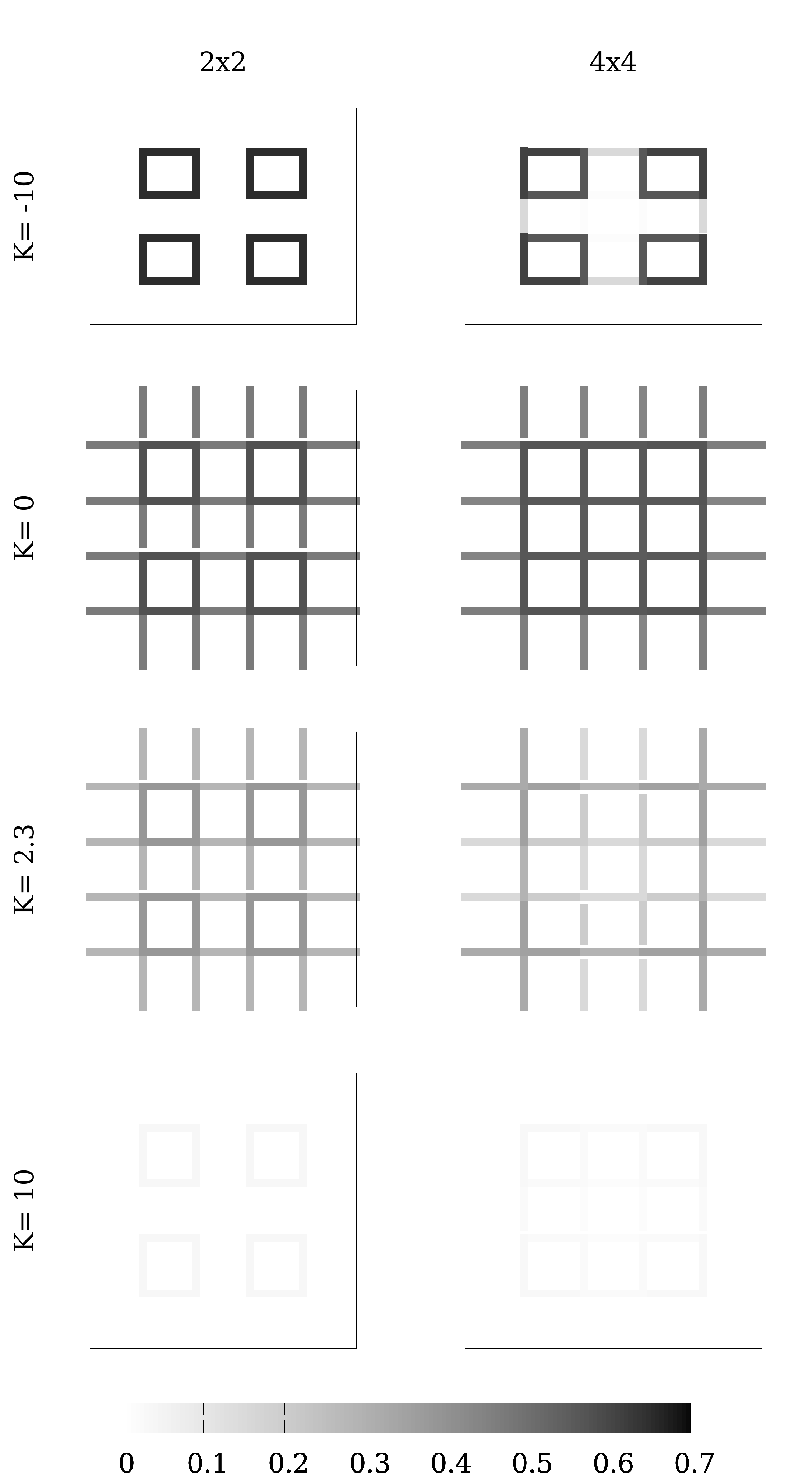}
\caption{\label{bondOP}
Expectation value of the hopping operator (\ref{B}) computed within HMFT--2$\times$2 (left) and HMFT--4$\times$4 (right) for several values of $K$ along the half filling line $(\mu=0)$: $K=-10$ (VBS), $K=0$ (SF), $K=2.3$ (CSF), $K=10$ (CVBS$_{1/2}$). The panels display the bonds of four $2\times 2$ clusters surrounded by the inter-cluster bonds (left) and the corresponding $4\times 4$ cluster bonds surrounded by the inter-cluster bonds (right).}
\end{figure}

\subsection{Phase diagram}
In Fig. \ref{HMFTdiag} the phase diagram obtained by means of HMFT $(L=2,4)$ is displayed for both the frustrated and unfrustrated regions of the $(K,\mu)$ plane together with several cuts for which we analyze in detail the phase transitions within HMFT--4$\times$4. Except for the tiny region in which the CVBS$_{5/8}$ phase emerges, the majority of the phase diagram is unveiled using 2$\times$2 clusters as the basic degree of freedom. As it is shown in Fig. \ref{plaqOP} and Fig. \ref{bondOP}, the use of 2$\times$2 clusters already permits us to correctly describe the essential features of all the phases, while HMFT--4$\times$4 includes minor quantitative corrections. In particular, we observe a uniform pattern of the plaquette and hopping expectation values (\ref{Bsquare})-(\ref{Pcross}) within the uniform SF and CSF phases, as well as the alternating plaquette pattern characteristic of the VBS and CVBS$_{1/2}$ as already described by the 2$\times$2 approximation. Notice that within the CVBS$_{1/2}$ phase, the expectation value of the hopping operator (\ref{Bsquare})-(\ref{Bparal}) is negligible over the whole system, while the expectation value of the bond-chiral operator (\ref{bo_clas}) has a plaquette pattern similar to the one displayed by the hopping operator within the VBS phase (not shown).

In the unfrustrated region ($K<0$), the phase diagram obtained by HMFT presents a significant improvement as compared to a standard single site mean-field (Sec.~\ref{ClasPhaseDiag}), where the classical solution was either uniform SF or FO. The HMFT allows for stabilization of the gapped VBS phase for large enough negative $K$, in qualitative agreement with QMC results \cite{qmc}. Interestingly, for $\mu=0$ the transition point is found at $K^{2\times 2}_c=-5.1$ and $K^{4\times 4}_c=-5.9$, showing a slow convergence to the QMC result, $K^{QMC}_c\simeq-7.9$. Although the HMFT is able to capture phases with $(\pi,\pi)$ CDW order \cite{isaevJQ}, we have not found any sign of long-range CDW order. In particular, we have computed the staggered magnetization OP (\ref{mstag}) obtaining $M_s=0$ over the whole diagram. Furthermore, both the VBS and CVBS$_{1/2}$ solutions are stable under the application of an external staggered magnetic field, or under the addition of a small repulsive density-density interaction term to the Hamiltonian (\ref{JK}). However, the normalized spin structure factor (\ref{ms}) for a 4$\times$4 cluster is in agreement with previous works \cite{ebl,ebl_ladder,qmc_hf}, and we observe stronger quantum CDW  fluctuations closer to the $K$-only limit, regardless of the sign of $K$. Note in passing that Sandvik and co-workers found using QMC simulations~\cite{qmc_hf,Sandvik06}, on the unfrustrated side of the phase diagram, a VBS-CDW transition at $K\simeq -14.5$.
%
\begin{figure}[t]
\includegraphics[clip=true,width=\columnwidth]{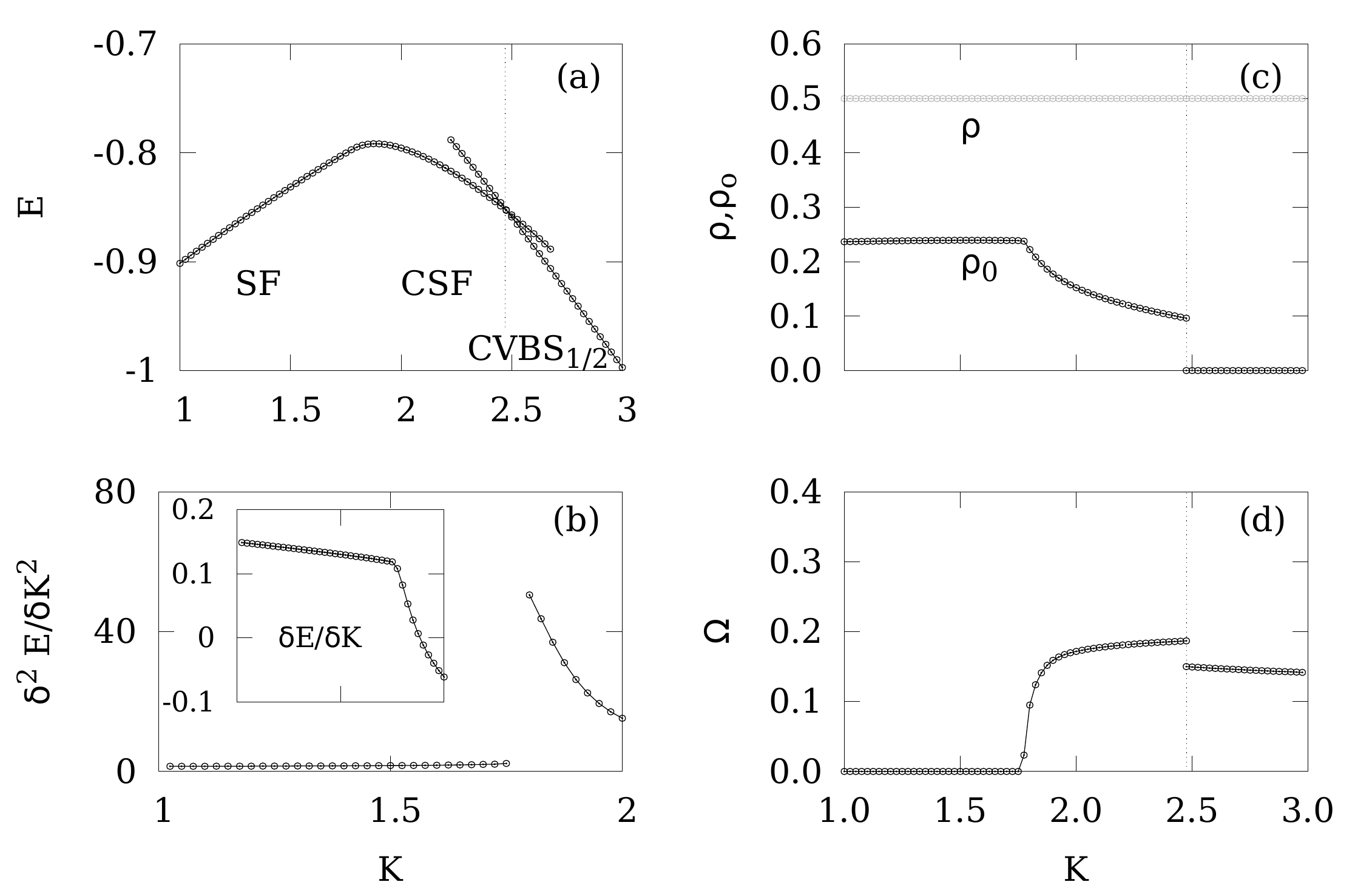}
\caption{\label{44cutA}(a) Energy for the SF-CSF and CSF-CVBS$_{1/2}$ phases at $\mu=0$ (cut $a$ in Fig. \ref{Diag_hartree}). (b) Second and first (inset) derivatives of the energy. (c) Total (gray) and condensate (black) densities. (d) Bond-chiral order. Dotted lines mark the first order transition between CSF and CVBS$_{1/2}$ phases. Solid lines are guides to the eye. The SF-CSF phase transition is  continuous, presumibly of second order.}
\end{figure}
%
\begin{figure}[t]
\includegraphics[clip=true,width=\columnwidth]{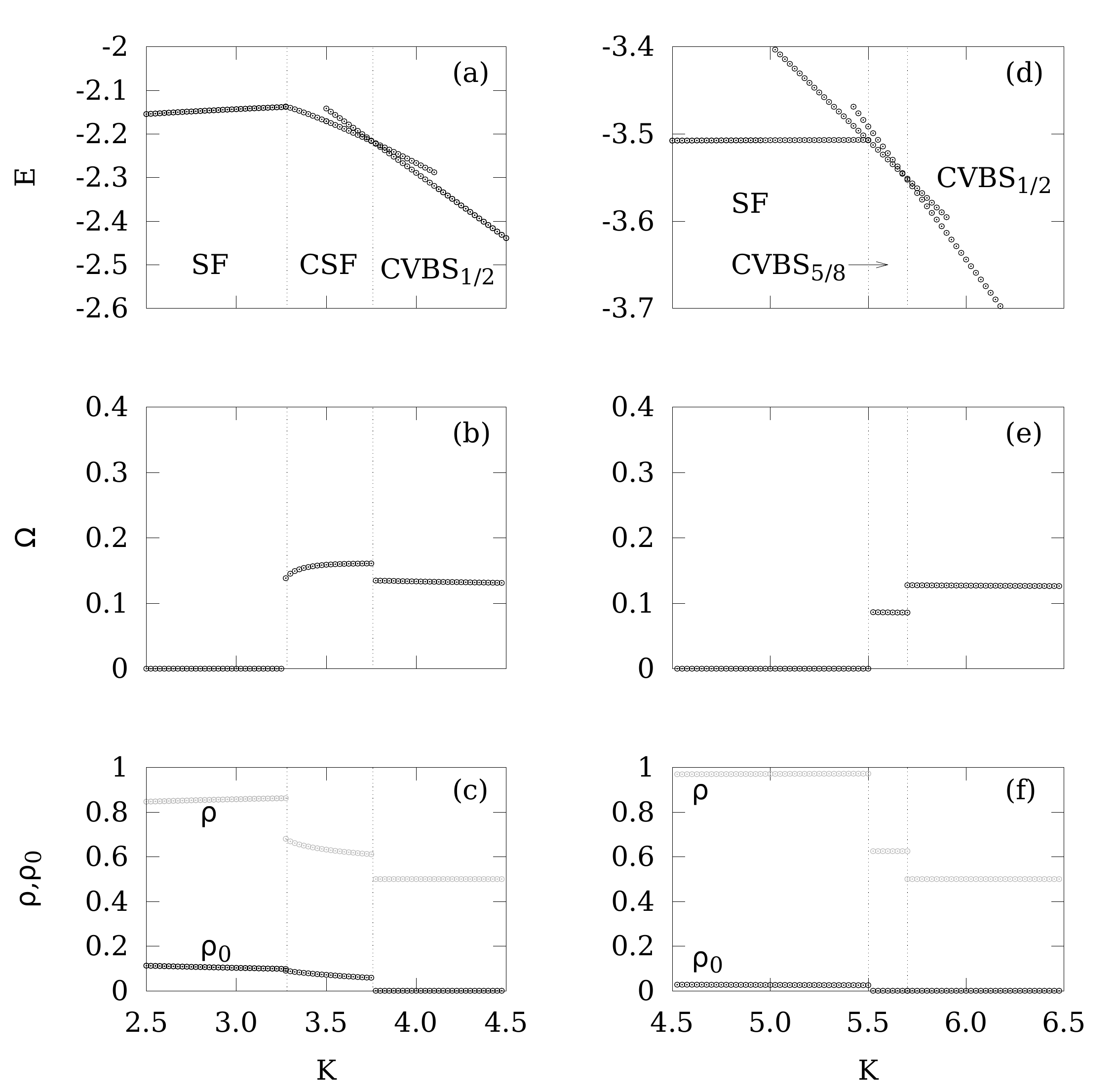}
\caption{\label{44cutBC}
Energy, bond-chiral order, and total and condensate densities for $\mu=2$ (cut $b$ in Fig. $\ref{HMFTdiag}$) are shown in panels (a), (b), and (c), respectively. The same observables, for $\mu=3.5$ (cut $c$ in Fig. $\ref{HMFTdiag}$) are shown in panels (d), (e), and (f). Dotted lines mark the first order phase transitions.}
\end{figure}
%
\begin{figure}[t]
\includegraphics[clip=true,width=\columnwidth]{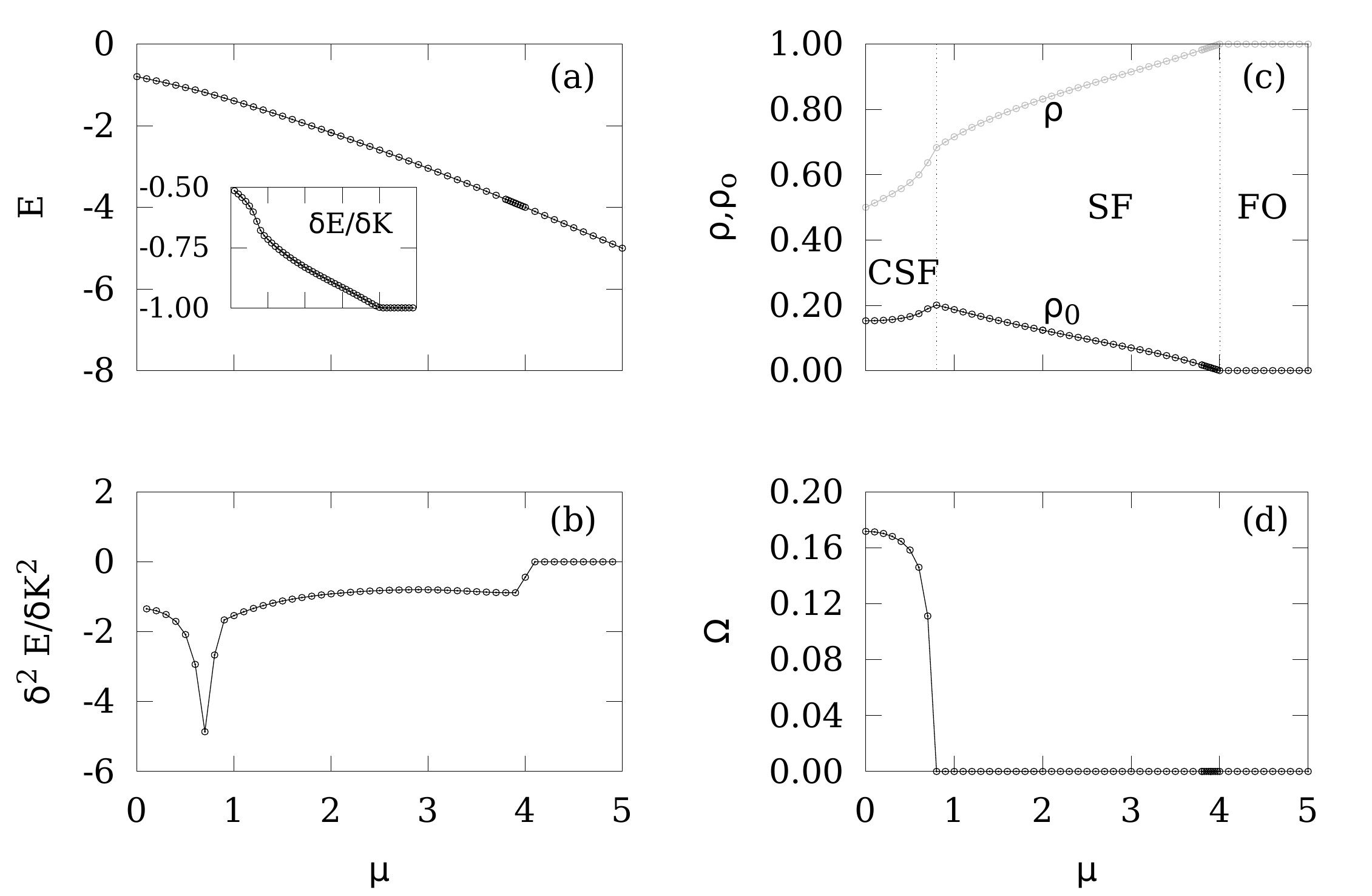}
\caption{\label{44cutD}(a) Energy and its first derivative (inset), (b) second-order derivative, (c) total and condensate densities, and (d) bond-chiral order for $K=2$ 
(cut $d$ in Fig. $\ref{HMFTdiag}$).}
\end{figure}

Figure $\ref{44cutA}$ displays the energy, total and condensate densities and the bond-chiral OP across the SF-CSF and CSF-CVBS$_{1/2}$ transitions at half filling $(\mu=0)$ in the frustrated regime $(K>0)$ (cut $a$ in Fig. \ref{HMFTdiag}). Also displayed are the first and second-order derivatives of the energy across the SF-CSF transition. The continuity of the order parameters and the derivatives of the energy across the SF-CSF transition suggest that it is of the second order type, while the jump of the order parameters and the energy crossing along the CSF-CVBS$_{1/2}$ transition indicates that it is of the first order type.

Figure $\ref{44cutBC}$ displays the energy, total and condensate densities and the bond-chiral OP across the SF-CSF-CVBS$_{1/2}$ and SF-CVBS$_{5/8}$-CVBS$_{1/2}$ transitions at $\mu=2$ (cut $b$ in Fig. \ref{HMFTdiag}) and $\mu=3.5$ (cut $c$ in Fig. \ref{HMFTdiag}), respectively. In all cases, the phase transitions are of the first order type, as they are signaled by discontinuities in the order parameters and the level-crossing of the energies. At $\mu\simeq1.5$, a potential TCP exists in the SF-CSF boundary.

Figure $\ref{44cutD}$ displays the CSF-SF and SF to FO transitions along $K=2$ (cut $d$ in Fig. \ref{HMFTdiag}). The two transitions are continuous, presumably of the second order type, as they are signaled by the energy derivatives and the continuous vanishing of the order parameters.

\begin{figure}[t]
\includegraphics[clip=true,width=\columnwidth]{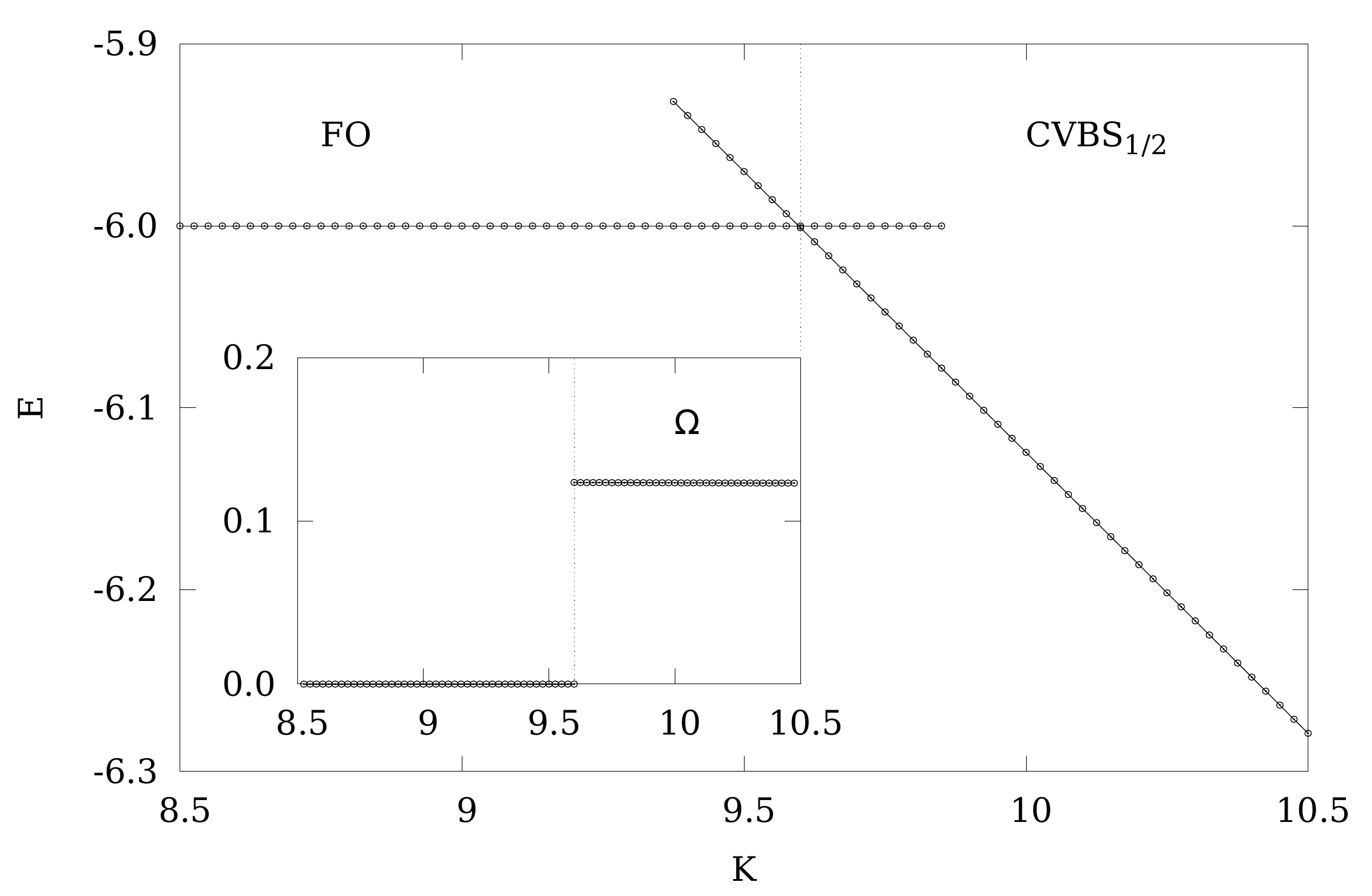}
\caption{\label{44cutE}
Energy and bond-chiral order parameter (inset) across the FO to CVBS$_{1/2}$ first order phase transition (cut $e$ in Fig. $\ref{HMFTdiag}$).}
\end{figure}

Figure $\ref{44cutE}$ displays the energy and the bond-chiral OP for the FO to CVBS$_{1/2}$ along $\mu=6$ for the frustrated regime $K>0$ (cut $e$ in Fig. \ref{HMFTdiag}). Both the crossing of the energy and the discontinuity of the order parameter indicates a first order phase transition.

%
%
\begin{figure}[t]
\includegraphics[clip=true,width=\columnwidth]{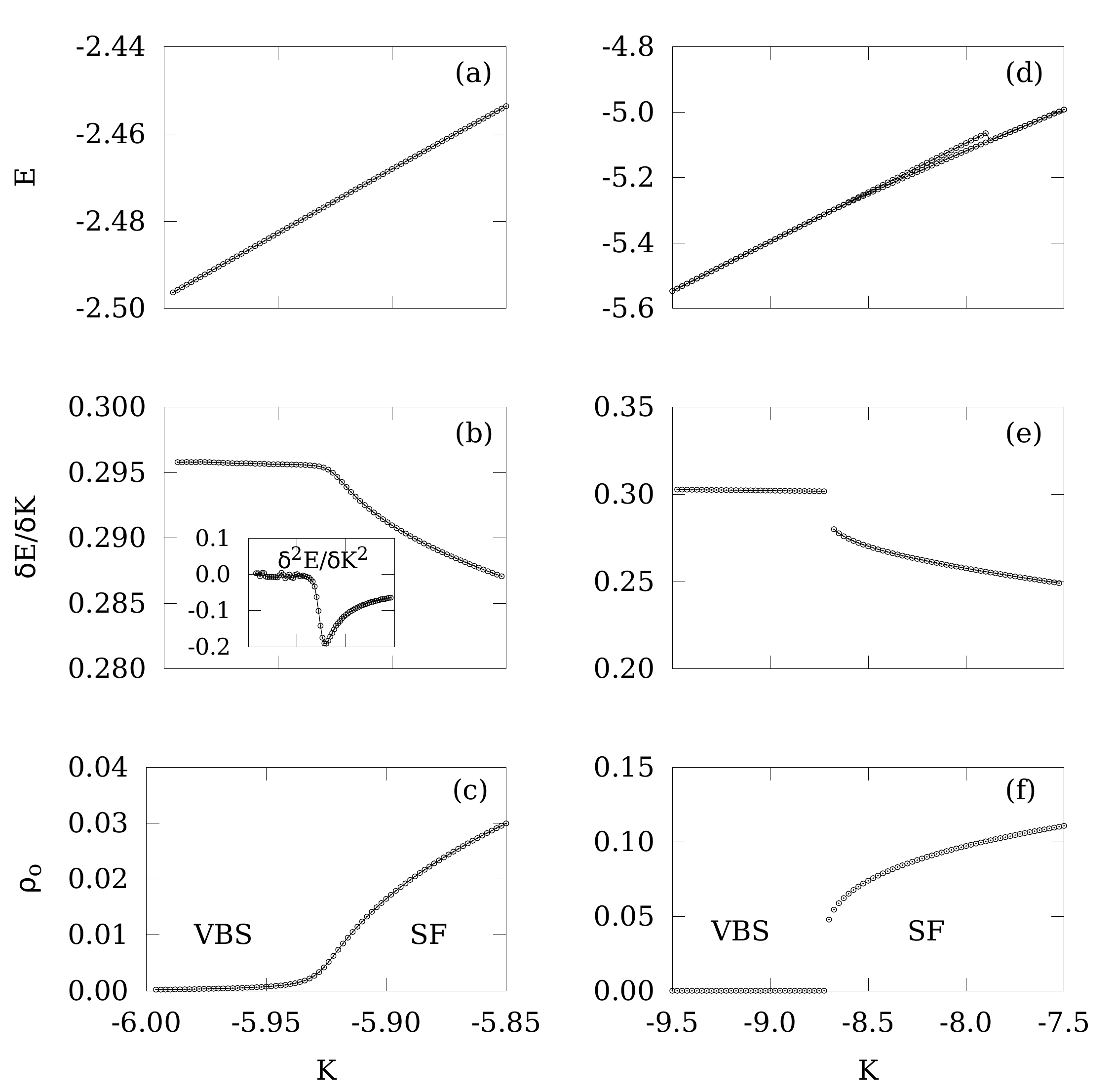}
\caption{\label{44cutFG}
(a) Energy, (b) its first and second-order  (inset) derivatives, and (c) condensate density across the VBS-SF transition at $\mu=0$ (cut $g$ in Fig. \ref{HMFTdiag}). The second derivative suggests that it is a continuous quantum phase transition, although we cannot discard the possibility of a weakly first order transition. (d) Energy, (e) its first derivative, and (f) condensate density across the VBS-SF transition at $\mu=4$ (cut $e$ in Fig. $\ref{HMFTdiag}$). It represents a first order phase transition, as it can be deduced from the discontinuity present both in the first derivative of the energy and in the condensate density order parameter.}
\end{figure}

Figure $\ref{44cutFG}$ displays the energy and its first and second-order derivatives for the SF-VBS transition at $\mu=0$ (cut $f$ in Fig. \ref{HMFTdiag}) and $\mu=4$ (cut $g$ in Fig. \ref{HMFTdiag}) for the unfrustrated regime $(K<0)$. In the first case, the continuous vanishing of the condensate density and the energy derivatives suggest a continuous phase transition. In this particular case, based on the cluster sizes used, we cannot definitively conclude whether the phase transition remains continuous or becomes weakly first order in the thermodynamic limit. In the second case, the first derivative of the energy and the vanishing of the condensate density suggest a first order phase transition. At approximately $\mu\simeq3.5$, a potential TCP exists, which separates the first and the second order phase transition along the VBS-SF boundary. 

\section{\label{conclusions}Summary and conclusions}
%
In this work we studied the quantum phase diagram of the $J$-$K$ model, for arbitrary densities, by means of the hierarchical mean-field theory (HMFT) \cite{hmft,isaevJQ}. This method is based on the identification of the main degrees of freedom providing the appropriate language that captures the relevant correlations of the quantum phases. 

Using $L$$\times$$L$ clusters of sizes $L=2,4$ as the new degrees of freedom, we have obtained a rich phase diagram where several superfluid and solid phases are characterized by emerging bond-chiral orders. Apart from the uniform superfluid and the trivial fully occupied  (empty) phases, we have encountered a bond-chiral superfluid and two novel valence bond-chiral solid phases characterized by an alternating expectation values of the plaquette and hopping operators along the $x$ and $y$ directions. Our main result is summarized in the phase diagram of 
Fig. \ref{HMFTdiag} with quantum phases schematically depicted in Fig. \ref{phases}.

We have shown how the use of clusters larger than a single site permits to unveil various solid phases which cannot be obtained by standard (single site) mean-field techniques. In particular, the classical approximation fails to correctly describe the ground state phase diagram of this model for ring-exchange intensities $K\geq \vert 2 \vert$. In the frustrated region, this approximation predicts a bond-chiral superfluid phase for $K>2$ which reduces to a tiny region when using HMFT--$L$$\times$$L$ $(L=2,4)$ giving rise to a new bond-chiral CVBS$_{1/2}$ phase. 

The phase diagram is mostly unveiled by means of HMFT--2$\times$2. The use of 4$\times$4 clusters includes minor quantitative corrections over HMFT--2$\times$2 results, with the exception of a tiny region of the CSF phase where a novel valence bond-chiral solid of density $\rho=5/8$, CVBS$_{5/8}$,  emerges. Although the limited size of the clusters may mask unusual phases characterized by correlations lengths greater than the ones comprised in a 4$\times$4 cluster, our results suggest that the structure of the phase diagram will remain in the thermodynamic limit. Computing with larger clusters (e.g. 6$\times$6, 8$\times$8,\ldots) might lead to the appearance of a mosaic of solid phases with commensurate densities in the narrow region mentioned above. Numerical studies with larger clusters would allow us to perform a rigorous finite-size scaling, however, this is highly demanding from a computational standpoint.

As the original and the cluster degrees of freedom are related by a canonical mapping, HMFT offers the possibility to compute low-lying excitations within a unified framework. In particular, being HMFT--1$\times$1 equivalent to the classical approximation, we have also shown that the method offers a convenient way to compute spin-wave dispersions over a multiple-sublattice classical ground state of a Hamiltonian with non-trivial interactions, such as the CSF ground state present in the $J$-$K$ Hamiltonian.

We have also computed the phase diagram in the unfrustrated regime obtaining results in qualitative agreement with previous QMC calculations\cite{qmc_hf,qmc,Sandvik06}. However, we have not found the $(\pi,\pi)$ CDW phase and its phase transition to VBS predicted by QMC, within any of the approximations (classical, HMFT--2$\times$2, HMFT--4$\times$4), even if all these approximations have been able to capture this kind of phase in several other models. This discrepancy could be related to an abnormal intrinsic correlation length greater than the dimensions of the 4$\times$4 cluster utilized in our HMFT.

\begin{acknowledgements}
DH would like to thank S. Pujari for interesting discussions and LPT (Toulouse) for hospitality. This work has been partially supported by the Spanish MINECO Grants FIS2012-34479, BES-2010-031607, EEBB-I-12-03677, EEBB-I-13-06139. NL is supported by the French ANR program ANR-11-IS04-005-01.
\end{acknowledgements}

\appendix
\section{\label{matrixel} CB Matrix elements}
%
In this Appendix we derive the form of the one-, two-, and four-body tensors of a general CB Hamiltonian $(\ref{ccbh})$. Let us start with the one-body tensor. As explained in Sec. \ref{HMFT}, one-body CB terms account for all the original interactions which act within a cluster labeled by the index $R$. Taking this into account, the explicit form of the one-body CB tensor is
\begin{widetext}
\begin{eqnarray}
\left(T_{R}\right)^{\alpha}_{\beta}&=&
-\mu\sum_{\mathbf{n}}\sum_{j\in R} n_{j} U^{\alpha\ast}_{R\mathbf{n}}U^{\beta}_{R\mathbf{n}}
+\sum_{\mathbf{n'}}\sum_{\langle ij\rangle \in R}\left( U^{\alpha\ast}_{R \lbrace 1_i,0_j\rbrace}U^{\beta}_{R\lbrace 0_i,1_j\rbrace} + 
U^{\alpha\ast}_{R \lbrace 0_i,1_j\rbrace}U^{\beta}_{R\lbrace 1_i,0_j\rbrace} \right)\notag\\
&&+K\sum_{\mathbf{n'}}\sum_{\langle ijkl\rangle\in R}\left( 
U^{\alpha\ast}_{R \lbrace 1_i,0_j,1_k,0_l\rbrace}U^{\beta}_{R\lbrace 0_i,1_j,0_k,1_l\rbrace} 
+U^{\alpha\ast}_{R \lbrace 1_i,0_j,1_k,0_l\rbrace}U^{\beta}_{R\lbrace 0_i,1_j,0_k,1_l\rbrace} 
\right)\label{T},
\end{eqnarray}
\end{widetext}
where we have used the notation $\lbrace 1_{i},0_{j}\rbrace\equiv \left( \ldots,1_{i},0_{j},\ldots\right)$ to label any cluster state $\mathbf{n}$ with the occupation of sites $i,j\in R$ fixed to $1$ and $0$, respectively. The sums $\sum_{\mathbf{n'}}$ run over all configurations of the remaining sites. In the same way, the two-body CB tensor is,
\begin{widetext}
\begin{eqnarray}
\left(V_{R_1R_2}\right)^{\alpha_1 \alpha_2}_{\beta_1 \beta_2} &=&
\sum_{\mathbf{n_1}',\mathbf{n_2}'}\sum_{\langle ij\rangle}
\left(\begin{array}{c}
U^{\alpha_1 \ast}_{R_1 \{0_i\}}U^{\alpha_2 \ast}_{R_2 \{1_j\}}  U^{\beta_1}_{R_1 \{1_i\}}U^{\beta_2}_{R_2 \{0_j\}}
+U^{\alpha_1\ast}_{R_1 \{1_i\}}U^{\alpha_2\ast}_{R_2 \{0_j\}}  U^{\beta_1}_{R_1 \{0_i\}}U^{\beta_2}_{R_2 \{1_j\}} \end{array}\right)\label{V}\\
&&+K\sum_{\mathbf{n_1}',\mathbf{n_2}'} \sum_{\langle ijkl\rangle}
\left(\begin{array}{c}U^{\alpha_1\ast}_{R_1 \lbrace 1_i,0_j\rbrace}U^{\alpha_2\ast}_{R_2 \lbrace 0_k,1_l\rbrace}  U^{\beta_1}_{R_1 \lbrace 0_i,1_j\rbrace} U^{\beta_2}_{R_2 \lbrace 1_k,0_l\rbrace}
+U^{\alpha_1\ast}_{R_1 \{0_i,1_j\}}U^{\alpha_2\ast}_{R_2 \{1_k,0_l\}}  U^{\beta_1}_{R_1 \{1_i,0_j\}} U^{\beta_2}_{R_2 \{0_k,1_l\}}
\end{array}\right),\notag
\end{eqnarray}
\end{widetext}
where in the first sum $i\in R_1$ and $j\in R_2$ and, in the second one, $i,j\in R_1$ and $k,l\in R_2$. Finally, the explicit form of the four-body tensor accounts for the double hopping of bosons from the corners of two next-nearest neighbour clusters to the corners of the two opposite diagonal clusters, as it is schematically represented in Fig. \ref{HMFT44scheme},
\begin{widetext}
\begin{eqnarray}
\left(W_{R_1R_2R_3R_4}\right)^{\alpha_1\alpha_2\alpha_3\alpha_4}_{\beta_1\beta_2\beta_3\beta_4} &=&
K\sum_{\langle ijkl\rangle}\sum_{\mathbf{n_{1}' n_{2}' n_{3}' n_{4}'}}(
U^{\alpha_1\ast}_{R_{1}\{1_i\}}U^{\alpha_2\ast}_{R_{2}\{0_j\}}U^{\alpha_3\ast}_{R_{3}\{1_k\}}U^{\alpha_4\ast}_{R_{4}\{0_l\}}  
U^{\beta_1}_{R_{1}\{0_i\}}U^{\beta_2}_{R_{2}\{1_j\}}U^{\beta_3}_{R_{3}\{0_k\}}U^{\beta_4}_{R_{4}\{1_l\}}\notag\\
&&~~~~~~~~~~~~~~~~~~~
+U^{\alpha_1\ast}_{R_{1}\{0_i\}}U^{\alpha_2\ast}_{R_{2}\{1_j\}}U^{\alpha_3\ast}_{R_{3}\{0_k\}}U^{\alpha_4\ast}_{R_{4}\{1_l\}}
U^{\beta_1}_{R_{1}\{1_i\}}U^{\beta_2}_{R_{2}\{0_j\}}U^{\beta_3}_{R_{3}\{1_k\}}U^{\beta_4}_{R_{4}\{0_l\}}).\label{W}
\end{eqnarray}
\end{widetext}

\section{\label{LSWT}Linear Spin-Wave theory via Schwinger bosons}
%
Linear Spin-Wave Theory (LSWT) is a semiclassical approach which takes into account the quantum fluctuations around the classical solution on the assumption that these are small compared to the expectation value of the spin and, therefore, the classical ground state is a good approximation to the quantum ground state. The general procedure  followed, when applying LSWT to a spin Hamiltonian, consists of the following steps:  $(i)$ rotate the spin basis at each site aligning the quantization axis with the classical spin, $(ii)$ perform a Holstein-Primakoff (HP) approximation in which the Hamiltonian is expanded in terms of HP canonical boson operators up to order $1/S$, and $(iii)$ diagonalize the resulting quadratic Hamiltonian by means of a Bogoliubov transformation. By this means, we automatically obtain the quantum corrections to the classical energy and the Bogoliubov eigenvalues provide the magnon dispersion relation. The $1/S$ correction to other thermodynamic quantities (total density, condensate density, etc) is automatically accounted for by taking derivatives of the corrected ground state energy with respect to the variational and physical parameters (chemical potential, transverse field, etc) and evaluating them at the zero-point.

However, if we were interested in quantities which cannot be directly derived from the ground state energy, i.e., expectation values of observables other than the Hamiltonian, a more subtle analysis has to be done. For a detailed discussion on how to correctly compute the $O(1/S)$ corrections to expectation values in the semiclassical approach, see Ref. [\onlinecite{coletta}]. This analysis goes beyond the scope of the present paper, as we will be only interested in the SW magnon dispersion relation and corrections to the classical energy. 

The general procedure described previously can be straightforwardly applied to the $J$-$K$ model when accounting for quantum fluctuations over the SF ground state. It becomes lengthy and tedious, however, when the ground state is the CSF. For this reason, we will work within the HMFT--$L$$\times$$ L$ framework described above, and show that it is exactly equivalent to the usual procedure, although more advantageous when treating Hamiltonians with complex many-body interacting terms. For the particular case of 1$\times$1, the CB mapping (\ref{cmap}) is equivalent to the Schwinger boson mapping in the bosonic language. 

First, let us express Hamiltonian $(\ref{JK})$ in terms of Schwinger bosons $\lbrace b_{j0}^{(\dag)},b_{j1}^{(\dag)}\rbrace$, which create (annihilate) an empty (0) or occupied (1) state at site $j$ of the original lattice,
\begin{eqnarray}
H &=& - \sum_{\left\langle i,j \right\rangle} \left( 
b^{\dagger}_{i1}b^{\dagger}_{j0}b_{i0}b_{j1} + h.c. 
\right)- \mu \sum_{j} b^{\dagger}_{j1}b_{j1}
\notag\\
&&+K\sum_{\langle ijkl \rangle}
\left(b^{\dagger}_{i0}b^{\dagger}_{j1}b^{\dagger}_{k0}b^{\dagger}_{l1}b_{i1}b_{j0}b_{k1}b_{l0} + h.c.\right)\notag\\
&&-\lambda \left(b^{\dag}_{j0}b_{j0}+b^{\dag}_{j1}b_{j1}-1\right),
\label{JKcb}
\end{eqnarray}
where we have added the physical constraint,  $\sum_{n=0,1}b_{jn}^{\dag}b_{jn}=1$, via a Lagrange multiplier $\lambda$, playing the role of an effective chemical potential. The relevant quantum fluctuations accounted for by the LSWT and which lead to low-lying excitations of the classical ground states reside in the space orthogonal to the one determined by the classical solution at each site of the lattice. Let us re-express Hamiltonian (\ref{JKcb}) in a new basis in which the ground state is enconded in one flavor $(\sf{g})$ and the orthogonal space in the other $(\sf{p})$. As seen before, the CSF has a two-sublattice structure where the azimuth angle of the pseudospin takes the values $\pm \phi$ depending on the sublattice. Therefore, the canonical transformation among the new CBs has to include this information about the ground state,
\begin{eqnarray}
b_{j\alpha}^{\dagger}&=&\sum_{n}U^{\alpha}_{n}~b_{jn}^{\dagger}~,~\text{for $j\in  A$},\label{rotA}\\
b_{j\alpha}^{\dagger}&=&\sum_{n}\left(U^{\alpha}_{n}\right)^{\ast}b_{jn}^{\dagger}~,~\text{for $j\in B$}\label{rotB},
\end{eqnarray}
where $\alpha$ takes just two values $\sf{g}$ and $\sf{p}$, and $n=0,1$. We know from Sec. \ref{ClasPhaseDiag} that its explicit form is
\begin{equation}
\hat{U}=
\left(
\begin{array}{cc}
\sin\left(\theta/2\right) e^{{\mathrm i}\frac{\phi}{2}} & -\cos\left( \theta/2\right) e^{{\mathrm i}\frac{\phi}{2}} \\
\cos\left( \theta/2\right) e^{-{\mathrm i}\frac{\phi}{2}} & \sin\left(\theta/2\right) e^{-{\mathrm i}\frac{\phi}{2}}
\end{array} \right)\label{U},
\end{equation}
where the first column $(U^{\sf{g}}_{n})$, accounts for the classical solution (\ref{prod_wf}) and the second column $(U^{\sf{p}}_{n})$ accounts for the orthogonal space. Applying transformations (\ref{rotA}) and (\ref{rotB}) to the Hamiltonian (\ref{JKcb}),
\begin{eqnarray}
H&=& -\mu\sum_{j\in A} T^{\alpha}_{\beta} 
\left( b^{\dagger}_{j\alpha}b_{j\beta} + b^{\dagger}_{j+\hat{\mathbf{x}},\alpha}b_{j+\hat{\mathbf{x}},\beta}\right)\nonumber\\
&& -\sum_{j\in A}\sum_{ \mathbf{u} }
V^{\alpha\beta}_{\alpha' \beta'}
b^{\dagger}_{j\alpha}b^{\dagger}_{j+\mathbf{u},\beta}b_{j\alpha'}b_{j+\mathbf{u},\beta'}\nonumber\\
&& +\frac{1}{2}K\sum_{j\in A}\sum_{\mathbf{u},\mathbf{v}} 
W^{\alpha\beta\gamma\delta}_{\alpha'\beta'\gamma'\delta'}\notag\\
&&~~~~\times 
b^{\dagger}_{j,\alpha}b^{\dagger}_{j+\mathbf{u},\beta}
b^{\dagger}_{j+\mathbf{u}+\mathbf{v},\gamma}b^{\dagger}_{j+\mathbf{v},\delta}\nonumber\\
&&~~~~\times 
b_{j,\alpha'} b_{j+\mathbf{u},\beta'}
b_{j+\mathbf{u}+\mathbf{v},\gamma'} b_{j+\mathbf{v},\delta'}\notag\\
&&-\lambda\sum_{j\in A}\left(b^{\dagger}_{j\alpha}b_{j\beta} + b^{\dagger}_{j+\hat{\mathbf{x}},\alpha}b_{j+\hat{\mathbf{x}},\beta} -2\right),
\label{cbh}
\end{eqnarray}
where $\hat{\mathbf{x}},\hat{\mathbf{y}}$ are unit vectors,
$\mathbf{u}$ involves a sum over $\pm\hat{\mathbf{x}},\pm\hat{\mathbf{y}}$ in the second line, and $\mathbf{u}$ 
$(\mathbf{v})$ a sum over $\pm\hat{\mathbf{x}}$ $(\pm\hat{\mathbf{y}})$ in the third line. The matrix elements 
$T^{\alpha}_{\beta}$, $V^{\alpha\beta}_{\alpha'\beta'}$ and $W^{\alpha\beta\gamma\delta}_{\alpha'\beta'\gamma'\delta'}$ contain all the information about the original Hamiltonian (\ref{JK}) and the classical ground state. These matrix elements are explicitly given by
\begin{eqnarray}
T^{\alpha}_{\beta}&=& \sum_{n} n U^{\alpha\ast}_{n}U^{\beta}_{n}=\sum_{n} n U^{\alpha}_{n}U^{\beta\ast}_{n},\\
V^{\alpha\beta}_{\alpha' \beta'} &=&  U^{\alpha\ast}_{0}U^{\beta}_{1}  U^{\alpha'}_{1}U^{\beta'\ast}_{0} \notag\\ 
&&+ U^{\alpha\ast}_{1}U^{\beta}_{0} U^{\alpha'}_{0}U^{\beta'\ast}_{1},\\
W^{\alpha\beta\gamma\delta}_{\alpha'\beta'\gamma'\delta'} &=&
 U^{\alpha\ast}_{1}U^{\beta}_{0}U^{\gamma\ast}_{1}U^{\delta}_{0}  U^{\alpha'}_{0}U^{\beta'\ast}_{1}U^{\gamma'}_{0}U^{\delta'\ast}_{1} 
 \nonumber\\
&+& U^{\alpha\ast}_{0}U^{\beta}_{1}U^{\gamma\ast}_{0}U^{\delta}_{1} U^{\alpha'}_{1}U^{\beta'\ast}_{0}U^{\gamma'}_{1}U^{\delta'\ast}_{0}.
\end{eqnarray}
By construction, they ``keep memory'' of the bipartite nature of the CSF ground state and they are therefore link-dependent. 

\subsection{Classical solution}
%
In this new basis, the CSF product wave function (\ref{prod_wf}) can be rewritten in a form similar to (\ref{HMFTwf}) 
($\lvert 0\rangle = \lvert \downarrow\rangle$), 
\begin{equation}
\vert\psi\rangle=\prod_{j} b^{\dag}_{j\sf{g}}\vert 0\rangle.
\end{equation}
The expectation value of (\ref{cbh}) with this wave function is the classical energy (\ref{e_clas}), which expressed in terms of the matrix elements $T,~ V~\text{and}~W$ can be written in the following compact form,
\begin{equation}
\mathcal{E} = 2M\left(
-\mu T^{\sf{g}}_{\sf{g}}-2JV^{\sf{gg}}_{\sf{gg}}+KW^{\sf{gggg}}_{\sf{gggg}}\right),\label{H0}
\end{equation}
where $M=N/2$ is the number of sites on sublattice $A$ (half of the original lattice). It will be computationally convenient to cast the variational equations in the Hartree matrix form (\ref{Diag_hartree}). For this purpose, we can rewrite the unitary transformation (\ref{U}) as 
\begin{equation}
\hat{U}= 
\left(\begin{array}{cc} x\left[z_{{\mathrm{r}}}+{\mathrm{i}} z_{\mathrm i}\right] & -y\left[z_{\mathrm r}+{\mathrm i}z_{\mathrm i}\right]\\ y\left[z_{\mathrm r}-{\mathrm i}z_{\mathrm i}\right] & x\left[z_{\mathrm r}-{\mathrm i}z_{\mathrm i}\right] \end{array}\right),\label{grot}
\end{equation}
where $x,y,z_{\mathrm r},z_{\mathrm i}\in \mathbb{R}$, and compute derivatives with respect to the variational parameters, $x,~y,~z_{\mathrm r},~\text{and}~z_{\mathrm i}$, under the unitarity constraint. We split the \textit{amplitude} $\left(x=\sin(\theta/2),~y=\cos(\theta/2)\right)$ and \textit{phase} $(z_{\mathrm r}+{\mathrm i}z_{\mathrm i}=e^{{\mathrm i}\phi/2})$ parts for computational convenience. The Hartree equation therefore reduces to two coupled matrix equations
\begin{eqnarray}
\hat{h}^{\theta}\mathbf{x}&=&\lambda \mathbf{x},\\
\hat{h}^{\phi}\mathbf{z}&=&\eta \mathbf{z},
\end{eqnarray}
where $\mathbf{x}=(x,y)$, $\mathbf{z}=(z_{\mathrm r},z_{\mathrm i})$, and $\eta$ is a Lagrange multiplier enforcing $z_{\mathrm r}^{2}+z_{\mathrm i}^{2}=1$. Note that $\lambda$ works as a chemical potential which fixes to unity the total density of the Schwinger boson system, while $\eta$ has no physical relevance. The explicit form of $\hat{h}^{\theta}$ is 
\begin{eqnarray}
\hat{h}^{\theta}=
\left(
\begin{array}{cc}
0 & h^{\theta}_{12}\\
h^{\theta}_{12} & -\mu
\end{array}
\right),
\end{eqnarray}
where $h^{\theta}_{12}=-4\left(xy\right)\cos(2\phi)+ 4K\left(xy\right)^{3}\cos(4\phi)$, while $\hat{h}^{\phi}$ is given by
\begin{eqnarray}
\hat{h}^{\phi}=
\left(
\begin{array}{cc}
h^{\phi}_{11} & h^{\phi}_{12}\\
h^{\phi}_{12} & h^{\phi}_{22}
\end{array}
\right) ,
\end{eqnarray}
with matrix elements 
\begin{eqnarray}
h^{\phi}_{11}&=&-4\left(xy\right)^{2}+4K\left(xy\right)^{4}z_{{\mathrm r}}^{2} , \\
h^{\phi}_{12}&=&-12K\left(xy\right)^{4}z_{\mathrm r}z_{\mathrm i} , \\
h^{\phi}_{22}&=& 4\left(xy\right)^{2}+4K\left(xy\right)^{4}z_{\mathrm i}^{2} .
\end{eqnarray}

\subsection{Holstein-Primakoff approximation}
%
To compute the LSWT corrections to the energy and find the magnon dispersion relations over each classical ground state, we apply the HP transformation to the bosonic Hamiltonian (\ref{cbh}). As we have already expressed it in terms of the ground state $(\sf{g})$ and its orthogonal space $(\sf{p})$, the HP transformation simply reads \cite{auerbach}
\begin{eqnarray}
b^{\dagger}_{j\sf{p}}b_{j\sf{p}} &=& b_{j}^{\dag}b_{j},\\
b^{\dagger}_{j\sf{g}}b_{j\sf{g}} &=& 1- b_{j}^{\dag}b_{j},\\
b^{\dagger}_{j\sf{g}}b_{j\sf{p}} &=& \sqrt{1- b_{j}^{\dag}b_{j}} ~b_{j}, \\
b_{j\sf{p}}^{\dag}b_{j\sf{g}} &=& b^{\dagger}_{j}\sqrt{1- b_{j}^{\dag}b_{j}}\label{HP}.
\end{eqnarray}
The HP bosons $\{ b_{j}^{\dag}, b_{j}\}$ obey standard canonical commutation relations. Assuming that the fluctuations over the classical ground state are small, one can expand the Hamiltonian up to terms quadratic in the HP bosons and then Fourier transform,
\begin{eqnarray}
b^{\dagger}_{\mathbf{r}_j}&=&\frac{1}{\sqrt{L^{2}/2}}\sum_{\mathbf{k}\in BZ}
e^{-\mathrm{i}\mathbf{k}.\mathbf{r}_j}~b^{\dagger}_{\mathbf{k}},\\
b^{\dagger}_{\mathbf{r}_j+\mathbf{u}}&=&\frac{1}{\sqrt{L^{2}/2}}\sum_{\mathbf{k}\in BZ}
e^{-\mathrm{i}\mathbf{k}.\left(\mathbf{r}_j+\mathbf{u}\right)}~\tilde{b}^{\dagger}_{\mathbf{k}},
\label{FT}
\end{eqnarray}
%
\begin{figure}[t]
\includegraphics[clip=true,width=0.95\columnwidth]{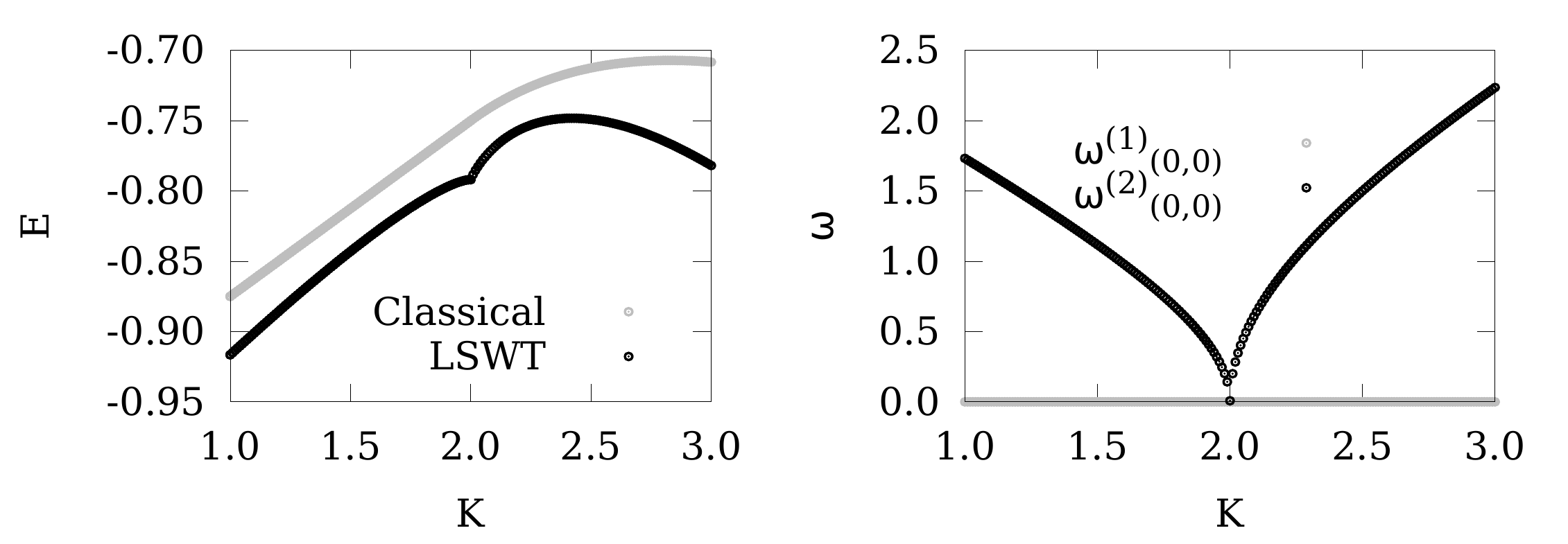}
\caption{\label{SWmu0}
Left: Classical energy (black) and LSWT energy (gray) for $\mu=0$ across the SF-CSF second order phase transition.
Right: $w^{(1)}_{(0,0)}$ (gray) and $w^{(2)}_{(0,0)}$ (black) Bogoliubov modes across the same transition.}
\end{figure}

\noindent
where we keep track of the two sublattices by adding a tilde for sublattice $B$. The first Brillouin zone ($BZ$), is defined as a square with vertices $\left( \pm\pi,0\right)$ and $\left( 0,\pm\pi\right)$. Finally, the Hamiltonian takes the form $H=\mathcal{E} + H^{(2)} +\ldots$, where the second order part of the Hamiltonian can be cast in matrix form \cite{ripka},
\begin{eqnarray}
H^{\left(2\right)}&=&
\sum_{\mathbf{k}}
\left( \beta_{\mathbf{k}}^{\dag},\beta_{\mathbf{-k}}\right)
\left(\begin{array}{cc}
\hat{A}_{\mathbf{k}} & \hat{B}_{\mathbf{k}}\\
\hat{B}_{\mathbf{k}}^{\ast} & \hat{A}_{\mathbf{k}}
\end{array}\right)
\left(\begin{array}{c}\beta_{\mathbf{k}}\\ \beta_{\mathbf{-k}}^{\dag}\end{array}\right)\notag\\
&&-\sum_{\mathbf{k}}\text{tr}\left(\hat{A}_{\mathbf{k}}\right)
\label{H2}
\end{eqnarray}
where we have defined $\beta^{(\dag)}_{\mathbf{k}}=(b^{(\dag)}_{\mathbf{k}},\tilde{b}^{(\dag)}_{\mathbf{k}})$ and $\hat{A}$ and $\hat{B}$ are $2\times2$ matrices with components
\begin{eqnarray}
A^{11}_{\mathbf{k}}=A^{22}_{\mathbf{k}}&=&-\frac{1}{2}\left( \mu T + \lambda \right)-2J V_{1}\notag\\
&&+2KW_{1}\left(1+\cos k_{x} \cos k_{y} \right) \\
A^{12}_{\mathbf{k}}=(A^{21}_{\mathbf{k}})^{\ast}&=& \left(-JV_{2}+2KW_{2}\right)\gamma_{\mathbf{k}} \\
B^{11}_{\mathbf{k}}=(B^{22}_{\mathbf{k}})^{\ast}&=& 2KW_{2}\cos k_{x}\cos k_{y} \\
B^{12}_{\mathbf{k}}=B^{21}_{\mathbf{k}}&=& \left(-J V_{1}+2K W_{1}\right)\gamma_{\mathbf{k}},
\end{eqnarray}
where $\gamma_{\mathbf{k}}= \cos k_x + \cos k_y$ and
\begin{eqnarray}
T&=& x^{2}\\
V_{1}&=&-2 y^2 x^2 \cos(2\phi) \\
V_{2}&=& x^4e^{{\mathrm i}2\phi}+ y^4e^{-{\mathrm i}2\phi} \\ 
W_{1}&=& -2 x^4 y^4 \cos(4\phi)\\
W_{2}&=& x^2 y^2 \left( x^4 e^{{\mathrm i}4\phi} + y^4 e^{-{\mathrm i}4\phi}\right).
\end{eqnarray}

Note that $ T,V_{1},W_{1}\in \mathbb{R}$ and $ V_{2},W_{2}\in \mathbb{C}$. Expression (\ref{H2}) can be diagonalized by a Bogoliubov transformation, $\beta^{\dag}_{\mathbf{k}}= X_{\mathbf{k}}\gamma^{\dag}_{\mathbf{k}}  + Y_{\mathbf{k}}\gamma_{\mathbf{-k}}$, leading to a Bogoliubov eigenvalue equation of the form\cite{ripka}, 
\begin{equation}
\left(\begin{array}{cc}\hat{A}_{\mathbf{k}}&\hat{B}_{\mathbf{k}}\\-\hat{B}_{\mathbf{k}}^{\ast}&-\hat{A}_{\mathbf{k}}^{\ast}\end{array}\right)
\left(\begin{array}{c}X_{\mathbf{k}}^{(n)}\\Y_{\mathbf{k}}^{(n)}\end{array}\right)=w_{\mathbf{k}}^{(n)}
\left(\begin{array}{c}X_{\mathbf{k}}^{(n)}\\Y_{\mathbf{k}}^{(n)}\end{array}\right),
\end{equation}
where SW dispersion relations are given by the two positive Bogoliubov eigenvalues, $w^{(2)}_{\mathbf{k}}\ge w^{(1)}_{\mathbf{k}} \ge 0$. Note that $X$ and $Y$ are $2\times 2$ matrices. The LSWT correction to the classical energy is simply,
\begin{equation}
E_{SW}= \mathcal{E} -\sum_{\mathbf{k}}\text{tr} \hat{A}_{\mathbf{k}}
+\sum_{\mathbf{k}}\left( w^{(1)}_{\mathbf{k}} + w^{(2)}_{\mathbf{k}} \right).
\end{equation}

In Fig. \ref{SWmu0} we show the computed SW correction to the energy for the SF-CSF transition at $\mu=0$. We observe that the transition becomes first order when adding the SW corrections, as we can distinguish a clear discontinuity in the first derivative of the energy. Note that the transition point is still placed at the very same value as it was in the classical approach, that is, at $K_c=2$. Both the SF and the CSF are gapless $(w^{(1)}_{(0,0)}=0)$ and have a finite value of the $w^{(2)}_{(0,0)}$ excitation mode, which vanishes continuously at the critical point. Note that within the uniform SF phase, this $w^{(2)}_{(0,0)}$ mode would correspond to the $(\pi,\pi)$ one-band SW dispersion mode, have not we performed a bipartition of the lattice.



\end{document}